\documentclass[9pt,sigconf,letterpaper]{acmart}

\acmYear{2021}\copyrightyear{2021}
\acmConference[IMC '21]{ACM Internet Measurement Conference}{November 2--4, 2021}{Virtual Event, USA}
\acmBooktitle{ACM Internet Measurement Conference (IMC '21), November 2--4, 2021, Virtual Event, USA}
\acmPrice{15.00}
\acmDOI{10.1145/3487552.3487829}
\acmISBN{978-1-4503-9129-0/21/11}

\usepackage{xspace}
\usepackage[nolist,nohyperlinks]{acronym}
\usepackage{xcolor}
\usepackage{url}
\usepackage{xurl}
\usepackage{subcaption}
\usepackage{enumitem}
\usepackage[noend]{algpseudocode}
\usepackage{algorithm}
\algrenewcommand\algorithmicrequire{\textbf{Input:}}
\algrenewcommand\algorithmicensure{\textbf{Output:}}
\usepackage{subcaption}
\usepackage{algorithmicx}
\usepackage{fancyvrb}
\usepackage{adjustbox}

\graphicspath{{figures/}}

\acrodefplural{AS}[ASes]{Autonomous Systems}
\acrodefplural{OS}[OSes]{Operating Systems}

\newcommand{\vfour}{IPv4\xspace}
\newcommand{\vsix}{IPv6\xspace}
\newcommand{\icmpsix}{ICMPv6\xspace}
\newcommand{\eui}{EUI-64\xspace}

\newcommand{\ehost}{endhost\xspace}
\newcommand{\ehosts}{endhosts\xspace}
\newcommand{\lhop}{last hop\xspace}

\newcommand{\yarrp}{yarrp\xspace}
\newcommand{\zmap}{zmap\xspace}
\newcommand{\eg}{e.g.,\xspace}
\newcommand{\ie}{i.e.,\xspace}
\newcommand{\etal}{et al.\xspace}

\newcommand{\punkt}[1]{\item\textbf{#1}:}

\begin{document}

\title{Follow the Scent: Defeating IPv6 Prefix Rotation Privacy}

\author{Erik Rye}
  \affiliation{
    \institution{CMAND}
  }
  \email{rye@cmand.org}

\author{Robert Beverly}
  \affiliation{
    \institution{Naval Postgraduate School}
  }
  \email{rbeverly@nps.edu}

\author{kc claffy}
  \affiliation{
    \institution{CAIDA/UC San Diego}
  }
  \email{kc@caida.org}

\begin{abstract}
\vsix's large address space allows ample freedom for choosing
and assigning addresses.  To improve client privacy and resist IP-based tracking, 
standardized techniques
leverage this large address space,
including privacy extensions and provider prefix rotation.
Ephemeral and dynamic \vsix addresses confound not only 
tracking and traffic correlation 
attempts, but also traditional network measurements, logging, and
defense mechanisms.
We show that the intended anti-tracking capability of these 
widely deployed mechanisms is unwittingly subverted by edge routers 
using legacy \vsix addressing schemes that embed unique identifiers.   

We develop measurement techniques that exploit these legacy
devices to 
make tracking such moving \vsix clients 
feasible by combining intelligent search space reduction with
modern high-speed active probing.
Via
an Internet-wide measurement campaign, we discover 
more than 9M affected edge routers and
approximately 13k /48
prefixes employing prefix rotation in hundreds of ASes worldwide.
We mount a six-week campaign to characterize the size and dynamics 
of these deployed \vsix rotation pools, and demonstrate via a 
case study the ability to remotely track client address movements 
over time.
We responsibly disclosed our findings to 
equipment manufacturers, at least one of which 
subsequently changed their default addressing logic.  
\end{abstract}



\maketitle

\section{Introduction}
\label{sec:intro}

\vsix deployment has seen rapid growth in recent years, with
residential broadband being one significant driver~\cite{akamaiblog}.
The large \vsix address space affords networks and clients considerable
flexibility in allocating and using addresses, as well as posing
unique ``needle-in-the-haystack'' challenges to network measurement.
For instance, best practices for \vsix dictate a /64 allocation or
larger~\cite{rfc6177} -- implying that a single residential customer
has more \vsix addresses than the entire \vfour address space.  Within
their allocated prefix, client addresses can change regularly as clients
choose random and ephemeral addresses via \vsix \emph{privacy
extensions}~\cite{rfc4941}, a standard to resist adversarial tracking
and correlation.  As we show in this work, some providers make use of the large \vsix address space
to further improve privacy by also, periodically and regularly, changing the entire prefix
allocated to customers~\cite{rfc8415} -- a process we term \emph{prefix
rotation}.

In this work, we develop techniques that allow us to mount an
Internet-wide active campaign to measure the prevalence of \vsix
prefix rotation.  At the heart of our approach is the surprisingly
common presence of deployed \ac{CPE}, \ie routers inside customer's
homes, that use a legacy addressing standard employing
\eui~\cite{rfc4862}.  In this form of \vsix addressing, the lower
64 bits of the CPE's address, the \ac{IID}, are unique, fixed, 
and persistent.
By sending active probes into candidate
network pools, we elicit \vsix responses containing the embedded IID
from these \ac{CPE}.  Our insight is to exploit these legacy static IIDs to
measure and characterize the real-world operational deployment of prefix rotation in terms
of provider \vsix prefix pool sizes, delegation sizes, and rotation
strategies.  

A second consequence of \ac{CPE} using static \eui IIDs is the potential to 
track clients as their prefix rotates.
The primary obstacle to finding a client's new prefix after a 
rotation is 
the very large \vsix search space;
it is impractical to probe the entirety of a provider's 
prefix, which is often as large as
a /32, \ie $2^{96}$.  To make tracking attacks feasible, we 
combine high-speed active probing~\cite{durumeric2013zmap, zmap6} and extend recent
innovations in \vsix topology discovery~\cite{edgy}.
With our novel technique, we 
demonstrate how a remote third-party can efficiently track 
a residential \vsix client with affected CPE, even as their entire address, including its
prefix, changes over time.  

\begin{figure*}[ht!]
 \centering
 \resizebox{1.7\columnwidth}{!}{\includegraphics[]{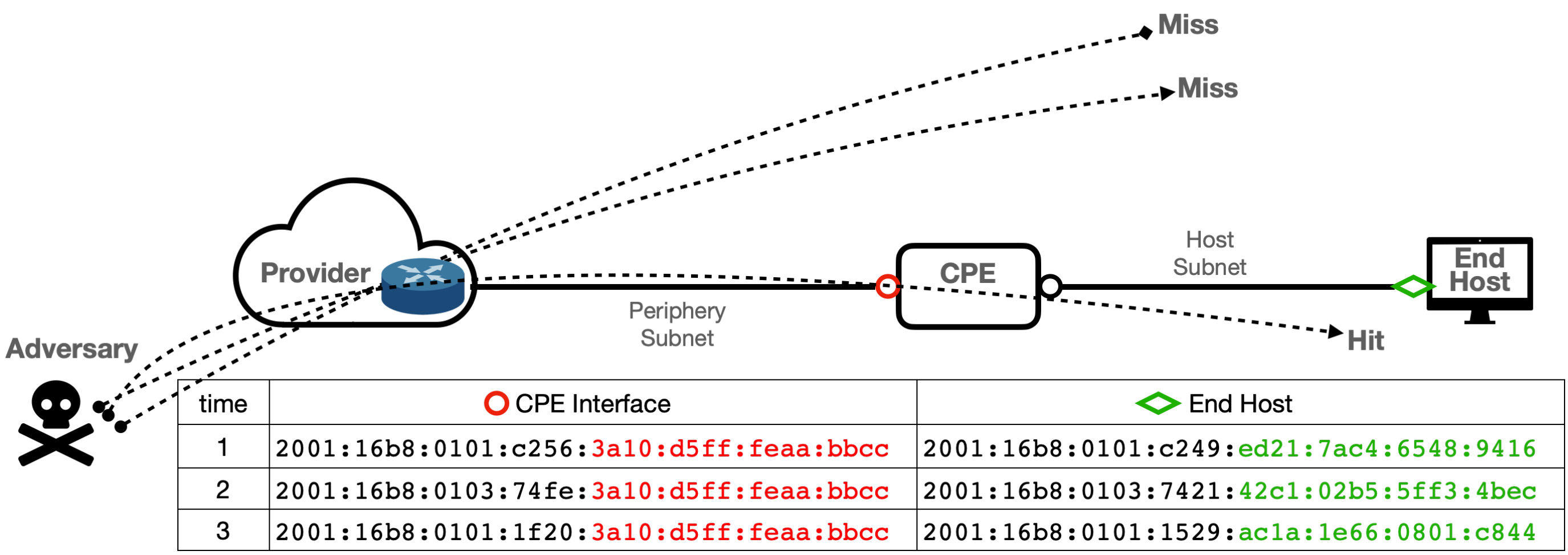}}
 \vspace{-2mm}
 \caption{In \vsix, the CPE (\eg home router) is a routed hop 
          and both the CPE and
          end host have public addresses.  The endhost's address
          is ephemeral: the lower 64 bit IID changes (privacy
          extensions) while the upper 64 bit prefix changes 
          (provider rotation).  Despite this, an attacker can
          track unsuspecting users when their home router uses
          a static lower 64 bits (in red).  We show how an
          adversary can efficiently send active probes to find
          a CPE, thereby revealing these users.}
 \label{fig:threat}
\end{figure*}

Next, we mount a large-scale Internet-wide active measurement campaign to characterize the scope 
and extent of \vsix prefix rotation in the wild by 
probing $\sim$18k /48
prefixes daily.  Our measurements discover 110M unique \eui addresses
assigned to CPE using 9M
distinct IIDs, \ie the same IID is seen in multiple prefixes -- 
demonstrating that IIDs are moving
between prefixes in practice.  Such prefix rotation is evident in
approximately 100 ASes across 25 different countries.  While these
confirmed rotators are a small fraction of all \vsix ASes, they
represent a lower-bound of at least 9M Internet users.

Not only does our work shed light on this previously understudied
aspect of production \vsix deployment, it highlights the fact that
a non-trivial fraction of deployed \ac{CPE} unintentionally subvert 
\vsix privacy mechanisms by permitting users and networks to
be tracked by proxy.  
In sum, our contributions
include:

\begin{itemize}
\addtolength{\itemsep}{-0.1\baselineskip}
 \item A measurement technique to identify providers that implement
       \vsix prefix 
       rotation, and large-scale measurements to discover
        and characterize these allocations (\S\ref{sec:methodology}).
 \item Demonstration that currently deployed \vsix anti-tracking and
       anti-correlation mechanisms (ephemeral addresses and prefix rotation) 
       can be rendered moot by CPE that implement legacy 
       standards (\S\ref{sec:methodology}).
 \item Internet-wide measurements showing that this problem
       affects $>$9M customers across dozens of networks
       worldwide (\S\ref{sec:campaign}).
 \item Per-network CPE homogeneity analysis, revealing
       that half of 87 studied networks contain a single manufacturer
       that dominates $\ge$80\% of their deployment
       (\S\ref{sec:results}).
 \item A case study demonstrating the ability
       to correlate seemingly unrelated 
       \vsix traffic flows over time using our methods with 60-90\%
       accuracy (\S\ref{sec:case}).
 \item Vulnerability remediation via communication
       with a major CPE vendor
       (\S\ref{sec:remediation}).
\end{itemize}

The remainder of this paper is organized as follows.
Section~\ref{sec:background} provides background on privacy-relevant
features and
standards in \vsix, and explores three different
security and privacy threats and harms enabled by CPE with \eui 
addressing.  Next, Section~\ref{sec:methodology} describes our active
measurement method to 
bound, per-provider, the space of possible locations to which a CPE and its
hosts may have moved.  We show that
this vulnerability affects millions of devices and customers in
Section~\ref{sec:campaign} and detail the vulnerable ecosystem in
Section~\ref{sec:results}.  Section~\ref{sec:case} demonstrates the real-world
feasibility of our approach in the
wild via a case study to track CPE and hosts over time. Section~\ref{sec:ethics}
describes the ethical guidelines adhered to during this study.
Finally, we detail our experience with an
equipment manufacturer in Section~\ref{sec:remediation} and conclude
with recommendations for long-term remediation.

\section{Background and Related Work}
\label{sec:background}

\subsection{Background}
Operational security and privacy issues of \vsix have been well
cataloged~\cite{rfc7721}.  Of particular relevance to our work
are three \vsix mechanisms: 1) SLAAC addresses; 2) privacy extensions; 
and 3) prefix rotation.

\begin{itemize}
 \punkt{SLAAC addresses} When using \eui \ac{SLAAC}
addresses~\cite{rfc4862}, an \ehost forms its full 128 bit \vsix
address by combining the provider-assigned prefix with 64 least-significant 
bits (the IID or ``interface identifier'') that are
derived from the interface's IEEE hardware MAC address.  This feature
ensures address uniqueness but exposes to layer-3 the host's layer-2
information, which allows the host to be trivially tracked across
network changes.  

 \punkt{Privacy extensions} \vsix clients commonly use \emph{privacy
extension} addressing, wherein the client chooses a random, ephemeral
lower 64-bit \ac{IID} \cite{rfc4941} in order to resist IP-based
tracking and correlation.  As per the RFC, ``changing the interface
   identifier over
   time makes it more difficult for eavesdroppers and other information
   collectors to identify when different addresses used in different
   transactions actually correspond to the same node.''
Thus, whereas multiple hosts behind a NAT
may map to one public \vfour address, a single residential host may
use many different public \vsix addresses.  Indeed, a large CDN found
that more than 90\% of \vsix addresses appear only once in a
long-running data collection~\cite{Plonka:2015:TSC:2815675.2815678}.  

 \punkt{Prefix rotation} While privacy extensions protect clients when
changing networks, IP-based tracking is still possible via the
customer's assigned prefix~\cite{rfc7707}.  Some providers
additionally deploy ``temporary-mode'' DHCPv6~\cite{rfc8415} in order
to regularly and periodically change the prefixes allocated to customers.
Such \emph{prefix rotation} is intended to prevent tracking -- now
both the IID (lower 64 bits) and the prefix (high 64 bits) change, often daily.

\end{itemize}

In \vsix, the \ac{CPE},
\eg the cable modem in a customer's home, is a routed hop.  
As shown in Figure~\ref{fig:threat}, 
between
the provider and the CPE is the \emph{periphery subnet}, while
a second \emph{host subnet} for the LAN 
may be part of the periphery subnet, or distinct.  
When probing toward any target address within a 
customer host subnet, active \vsix topology discovery elicits a
response from the CPE, the red colored interface in
Figure~\ref{fig:threat}.  

Whereas \vsix privacy extensions are enabled by default and used today in all
modern desktop and mobile operating systems, \eui
addresses are still commonly used in deployed \ac{CPE},
likely because many CPE run embedded, old,
unmaintained,
or
proprietary
operating systems.  
For example, Figure~\ref{fig:threat} shows 
the CPE obtaining new prefixes over three points in time.
Note that while the addresses remain in
the provider's allocated prefix (\texttt{2001:16b8::/32}), the periphery
prefix within the rotation pool changes for each sample and the
\eui IID remains constant (in this example, corresponding to the CPE's MAC
\texttt{38:10:d5:aa:bb:cc}).

\subsection{Related Work}

Recent large-scale \vsix traceroute measurements
discovered 30M \eui \lhop addresses in the Internet~\cite{edgy}
corresponding to \ac{CPE} interfaces facing the
upstream provider.  
While Rye's study found 30M distinct \vsix addresses with \eui IIDs,
only 16M of those IIDs were unique, implying that the periphery subnet
was changing and using prefix rotation~\cite{edgy}.  
While~\cite{edgy} developed an algorithm for general \vsix periphery discovery, this work explores how the
widespread use of legacy \eui addresses by CPE effectively removes \ehost
privacy enhancements offered by prefix rotation and ephemeral 
addresses. Other active measurement studies have explored \vsix topology
discovery more generally, rather than exclusively on periphery discovery, such
as Beverly \etal~\cite{imc18beholder} and Gasser
\etal~\cite{Gasser:2018:CEU:3278532.3278564}.

Related to our study are efforts to perform address-based blocking of 
infected or abusive \vsix hosts.  The same privacy-enhancing mechanisms designed to protect
users serve to complicate efforts
by \eg content providers, to block, filter, or rate-limit attacks by
\vsix address or prefix~\cite{imc20facebook}.  Our work sheds light
on the practical difficulties of protecting \vsix networks from attack, 
provides new insights on address lifetimes induced by prefix
rotation, and provides a potential means to track and correlate 
attack traffic as part of a defensive mechanism. Although recent
work~\cite{padmanabhan2020dynamips} examines \vsix address lifetimes using
several thousand privileged vantage points located within customer
networks~\cite{atlas}, we demonstrate that an unprivileged attacker can
infer the same information for orders of magnitude more networks using active
probing.

\section{Tracking Endhost Subnets}
\label{sec:methodology}

Discovering the instantaneous binding between a particular, known
\vsix \ehost and its CPE is straightforward: simply run a traceroute
toward that destination \vsix address to induce the CPE to return an
\icmpsix hop limit (n\'{e}e TTL) exceeded message.  Since the source \vsix address of this
\icmpsix message is the interface on the route toward the original
probe~\cite{rfc4443}, the response reveals the CPE's WAN address, which may
contain an \eui \ac{IID}. Tracking a user
by her CPE's static \eui \ac{IID}, after both her
periphery prefix and IID change, requires discovering
an \vsix traceroute destination that responds with this CPE's
\eui \ac{IID}.  \emph{Finding} the correct destination that will reveal the
binding, amid the vast \vsix address
space, is hard and the crux of the problem we tackle.

At first blush, this search space appears impractically
large: the lower 64 bits are all changed and the new subnet may
be anywhere in the provider's global prefix.  Since many providers
have an allocated (from their Regional Internet Registry) 
address prefix of /32 or larger, a
brute force search would require $2^{128-32}=2^{96}$ probes -- a 
size impossible to probe. We make this probing 
feasible via high-speed active
probing (\S\ref{sec:activeprobing}) and intelligent search space reduction
(\S\ref{sec:bounding}). 

Our methodology has two phases:
1) determining the size of customer prefix allocation for each provider
(\S\ref{sec:allocsize}); and
2) finding the rotation pool to which the customer belongs
(\S\ref{sec:rotationpool}). %
We describe our active probing technique to inferentially
perform each of these tasks next.

\subsection{Active Probing}
\label{sec:activeprobing}

Prior work, \eg \cite{edgy,imc18beholder}, utilized \yarrp
\cite{imc16yarrp} to perform
high-speed active \vsix topology mapping.  While \yarrp is ideal for full
topology probing (\ie finding the sequence of router interfaces along
a path to a destination), we are not interested in the entire
forwarding path, only the \lhop toward an \vsix destination. The
\acp{IID} of hosts within the customer prefix are ephemeral and
unknown.
However, the recommended \vsix behavior is for the \lhop to
return an \icmpsix unreachable 
message when an address probed at random within 
the prefix does not exist~\cite{rfc4443}.  
This \icmpsix message exposes
the CPE's address facing the provider, and requires only a single probe into the entire customer
prefix allocation.  That is, we need not select a responsive \ehost in
order to induce a response from the CPE responsible for the host
subnet.

Throughout this work, we utilize the zmap6 \vsix extensions~\cite{zmap6} to the high-speed \zmap
prober~\cite{durumeric2013zmap} developed by Gasser \etal 
We identify networks that exist internal to the
\ac{CPE} (the ``host subnet'' in Figure~\ref{fig:threat}), and
send \icmpsix Echo Request probes to random \acp{IID} in these host subnets. 
We probe at 10k packets per second from a well-connected vantage point
in a European IXP.

Sending probes to the host subnet typically generates
one of several \icmpsix messages. Generally, these responses are
\emph{Destination Unreachable} errors (\emph{Administratively
Prohibited}, \emph{No Route to Destination}, and \emph{Address Unreachable} are
common), but we also observe \emph{Hop Limit Exceeded} responses as well.
These type and code combinations indicate different 
OS
behaviors when receiving a packet destined for a nonexistent host in the
internal subnet.
The particular
response type itself does not matter in our application; all of 
these messages 
reveal the source address of the CPE in reply. Mistaking responses from intermediate hops
as CPE is unlikely: managed network infrastructure is typically
statically addressed, while we focus specifically on \eui \vsix addresses. Furthermore, the
\ac{MAC} addresses embedded in the \eui addresses we discover align with major CPE
vendors' \acp{OUI}.
\yarrp may elicit a
response in some instances that \zmap does not, \eg because of
particular \icmpsix filtering policies.
However, \zmap more unambiguously reveals the \vsix periphery because intermediate hops
do not reply with \emph{Hop Limit Exceeded} \icmpsix messages as they do in \yarrp.
Reduced \icmpsix error messages also
helps avoid potential \icmpsix response rate limiting~\cite{imc18beholder}.
Because we are interested
only in periphery (CPE) responses, as opposed to responses from core and access
infrastructure, using \zmap allows us to probe at a higher rate than
\yarrp without
receiving responses from portions of the Internet topology tangential to our
study. 

\subsection{Bounding the Search Space}
\label{sec:bounding}

\begin{figure}[t]
 \centering
 \includegraphics[width=\linewidth]{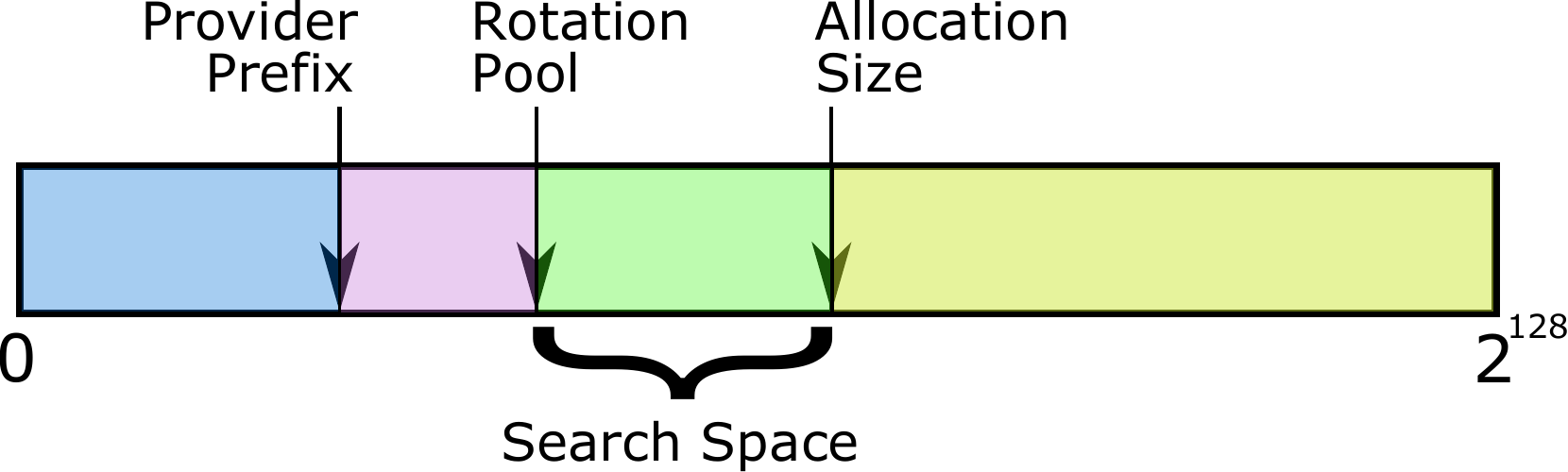}
 \caption{Limiting the search space to track \vsix hosts:
 for a provider, we infer the: i) 
 size of the allocations to customers; and ii) 
 range of prefixes used for rotation.}
 \label{fig:space}
\end{figure}

Bounding the search space requires understanding the size of prefix
allocated to customers (for instance a /64 versus a /56), the size of
the rotation pool (for instance bits 33-64 in
Figure~\ref{fig:threat}), and potentially the rotation pool
identifier (for providers that have more than one address 
pool).  A key challenge is that there is no standard or
operational consistency in \vsix deployment, hence each of these
variables \emph{differs by provider}.  A significant
contribution of our work is a large-scale Internet measurement
campaign to empirically characterize provider \vsix 
deployments and allocation behaviors
in the wild.

As depicted in Figure~\ref{fig:space}, the size of prefix that
a provider allocates to customers bounds the search space from above,
while the rotation pool size bounds the search space from
below.  Thus, rather than searching an impossibly large space, we need
only probe within the green shaded area to scalably locate a
particular CPE.  

In our canonical example (Figure~\ref{fig:threat}), both the CPE
interface and \ehost interface change over three time steps.
The adversary knows that a binding exists between the \ehost
and the MAC address of their CPE.  Here, the search space is reduced
to a rotation pool of a /46, while the customer prefix allocation size
is a /64.  The adversary thus probes random targets for each /64 within this 
/46 range
until it elicits a response from a CPE containing the correct \eui
\ac{IID} ($E[] = 2^{18-1}$ probes, or about 13 seconds at 10kpps).  
This example depicts two probes that pass through the 
provider's network but miss the CPE, while a third probe reaches
the CPE and therefore returns a response that reveals the CPE's new 
address to the attacker.

\begin{figure*}[t]
  \begin{subfigure}{.32\textwidth}
    \centering
    \includegraphics[width=.9\columnwidth]{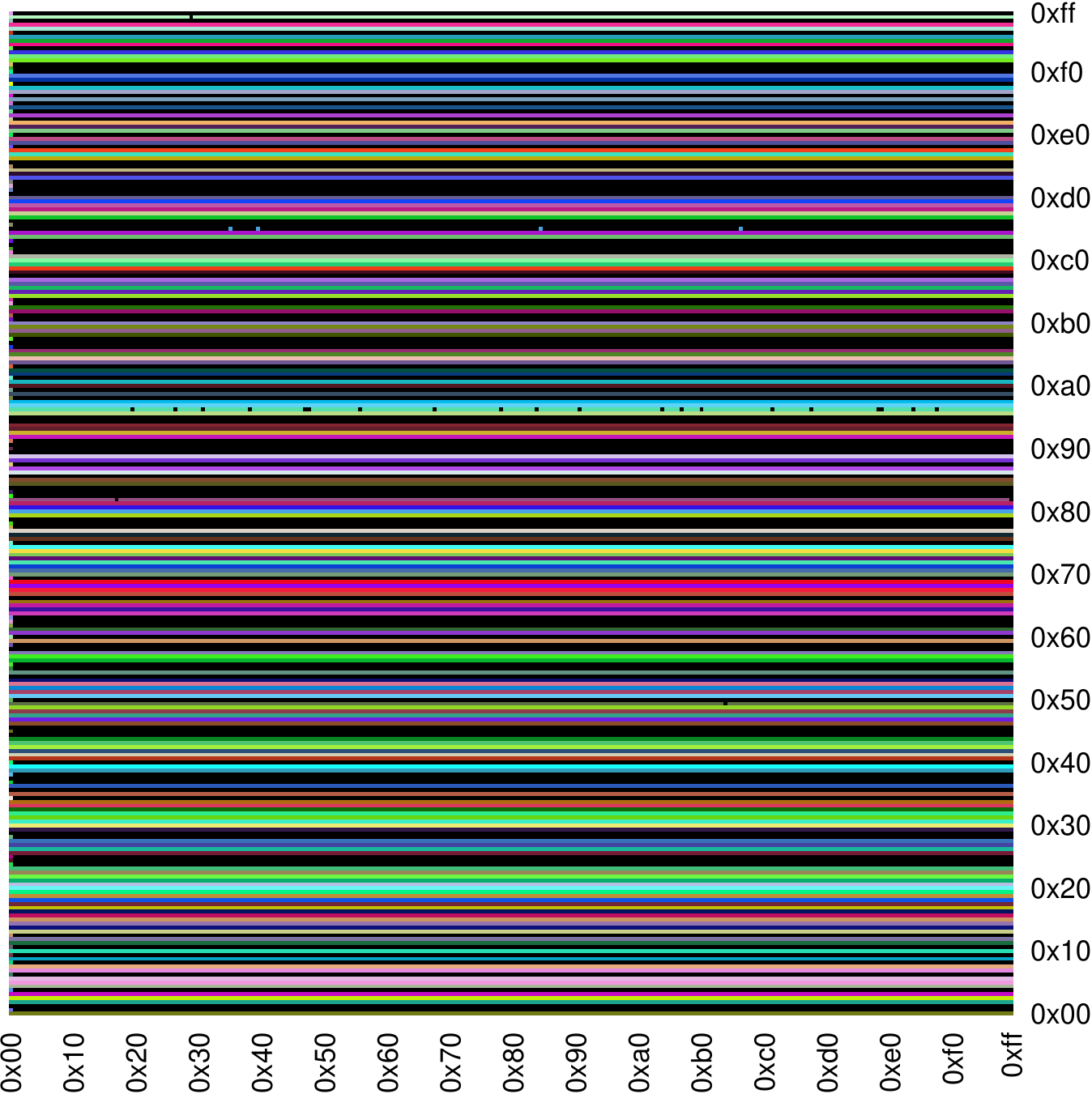}
    \caption{Entel (Bolivia): /56 allocations}
    \label{fig:fiftysix}
  \end{subfigure}%
  \begin{subfigure}{.32\textwidth}
    \centering
    \includegraphics[width=.9\columnwidth]{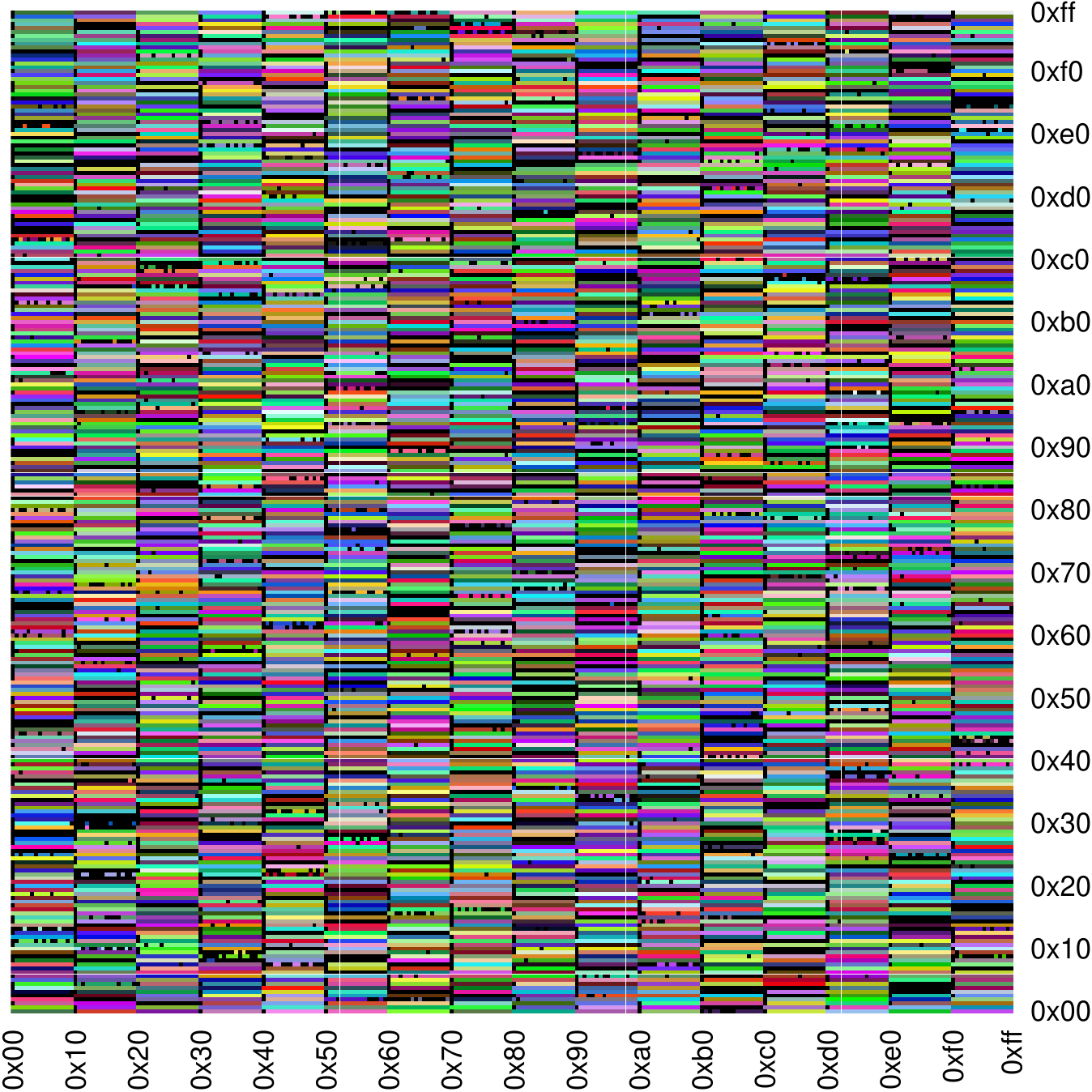}
    \caption{BH Telecom (Bosnia): /60 allocations}
    \label{fig:sixty}
  \end{subfigure}%
  \begin{subfigure}{.32\textwidth}
    \centering
    \includegraphics[width=.9\columnwidth]{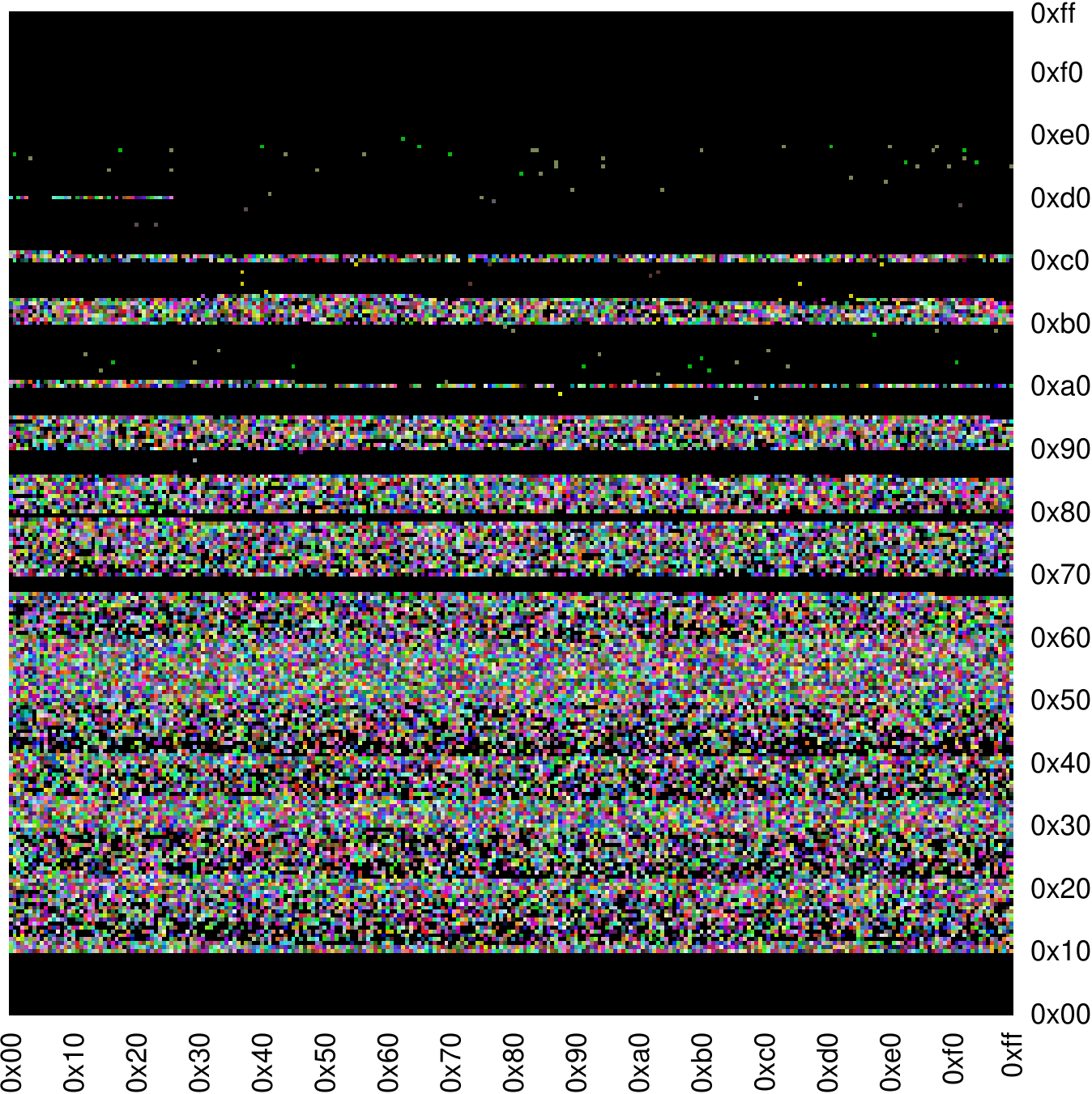}
    \caption{Starcat (Japan): /64 allocations}
    \label{fig:sixtyfour}
  \end{subfigure}%

  \caption{An adversary can infer providers' customer address allocation
  policies, enabling efficient scanning for targeted tracking
  (\S\ref{sec:case}). The $y$-axis of plots represents the $7^{th}$ byte of a
  probed address, while the $x$-axis denotes the $8^{th}$ byte; each pixel
  represents a probed /64 network. Each color represents a different responsive
  source address, while black indicates no response was received when probing to an
  address in that /64 network. }
  \label{fig:allocation_strategies}
\end{figure*}

\subsubsection{Customer Prefix Allocation Size Discovery}
\label{sec:allocsize}

We seek to better
understand \emph{how} providers allocate address space to their customers. 
RFC 6177~\cite{rfc6177} recommends that providers allocate to customers prefixes 
ranging in size from /48 to /64, but leaves the decision 
up to individual service providers. Because we target networks
\emph{internal} to \ac{CPE}, when our probes are unable to be routed by the
\ac{CPE} because of administrative policy or because the host to which they are
addressed does not exist, the \ac{CPE} replies with an \icmpsix error
that identifies it as responsible for that network 
(\S\ref{sec:activeprobing}).

Because \eui \ac{SLAAC} addressing requires a subnet no smaller than /64,
/64 networks are the smallest recommended allocation size~\cite{rfc3177,rfc6177}
for end sites.
Thus, when we probe a target within each constituent /64 subnet of
a rotation pool,
we infer the size of the host subnet 
internal to the \ac{CPE} by observing contiguous probed /64 networks that elicit
a response from the same address.

For example, consider a /48 network from which a provider allocates 
/56 subnets
to customers.
A \ac{CPE} device, if responsive to probes
addressed to destinations in its host subnet, will originate \icmpsix error
messages if those destinations cannot be reached. Many devices will
prohibit traffic to internal addresses that was not initiated from
the host 
subnetwork, and it is exceptionally unlikely to guess a
device's address behind the \ac{CPE}; therefore, in practically
all cases the probe traffic will be undeliverable. Thus, for each of
the $2^{64-56} = 256$
probes with random \acp{IID} sent to each /64 within the /56 assigned to the
\ac{CPE} device, the \ac{CPE} will originate an error message with its external
\vsix address (red interface in Figure~\ref{fig:threat}) as the source. By observing consistent source addresses in replies
for contiguous ranges of target addresses, we are able to infer how the provider
partitions its address space for customer prefix assignments.

To illustrate the methodology, we characterize three
different providers via active probing on September 7, 2020. 
Figure~\ref{fig:allocation_strategies} exposes three distinct 
allocation strategies deployed in practice. 
Figure~\ref{fig:fiftysix} represents a /48 prefix
belonging to Entel~\cite{entel}, a Bolivian telecommunications 
company. Each
horizontal colored band represents a different address responding to our 
probes. Colors are selected only to allow for visual
differentiation and to show contiguous regions; the particular color
is not significant in these figures.
The $y$-axis represents the $7^{th}$ byte of the probe destination address, and
$x$-axis represents the $8^{th}$ byte of the destination address, \ie
the last two octets of the high 64 bits. 

The consistent
response address for targets with a fixed $7^{th}$ byte, while varying the
$8^{th}$ byte (shown by the color banding), 
indicates that the provider in Figure~\ref{fig:fiftysix} allocates /56 prefixes.
The black horizontal bands, interspersed throughout the /48, are /56 networks
for which we received no response. These /56s may be unallocated by the
provider, or the devices assigned to those /56s may silently drop
our probes.

Figure~\ref{fig:sixty} shows the same analysis
from our probing of a Bosnian service provider's /48 (BH
Telecom~\cite{bihnet}).
BH Telecom allocates /60
networks to its customers, which are depicted as 16 short, horizontal lines
within each /56.  Some /60s are unallocated or the associated \ac{CPE}
is unresponsive to our probes, as evidenced by the contiguous black lines
interspersed throughout. Some /60s exhibit black pixels within them, indicating
potential packet loss or the \ac{CPE} device not responding to probes.

Finally, Figure~\ref{fig:sixtyfour} is a /48 prefix delegated to
a third provider, Starcat~\cite{starcat}, a
Japanese ISP. The customer-allocated subnets within this
/48 are /64s, depicted by the heavily-pixelated portions of the Figure. Of the
three sub-figures in Figure~\ref{fig:allocation_strategies},
Figure~\ref{fig:sixtyfour} exhibits the largest unresponsive portions, with
significant space in the upper quarter of the /48 prefix that did not respond to
our probes, likely because it was unallocated to any \ac{CPE}.

Algorithm~\ref{alg:allocPrefixSize} provides pseudocode
of our technique to infer an AS's prefix allocation sizes.
For each \ac{AS}, we
calculate the median inferred allocation size from all the \eui \acp{IID}'
inferred allocation sizes. 
Service providers may have one or more prefix allocation
sizes; providers that offer different classes of service or differentiate
between customer types exhibit distributions with clusters around multiple
prefix sizes. 

Inferring the customer prefix allocation size allows an adversary
to reduce the number of probes necessary to enumerate the \ac{CPE}
devices
present within a prefix. For instance, rather than sending a probe to an address
in each /64 within the Entel /48 in Figure~\ref{fig:fiftysix}, a tracker that
knows a priori that Entel allocates /56 networks to customers need only
send 256 probes (one to an address in each /56) in order to elicit a response
from each \ac{CPE}, decreasing probing cost by 99.6\%. In contrast, the Starcat
/48 in Figure~\ref{fig:sixtyfour} requires an adversary to probe each /64, as
/64s are allocated to \ac{CPE} devices.

\subsubsection{Identifying Rotation Pools}
\label{sec:rotationpool}

Next, we seek to determine the size of the rotation pools within which allocated customer
prefixes rotate per provider.  
Across all \eui addresses discovered in a provider, we 
first find the maximum numeric distance
between any two /64 periphery prefixes containing that \eui.
We then compute the median across all per-\eui ranges to determine
the likely pool size for the provider as a whole.
Pseudocode for this inference is provided in 
Algorithm~\ref{alg:rotationPoolSize} in the Appendix.
Algorithm~\ref{alg:rotationPoolSize} is essentially identical to
Algorithm~\ref{alg:allocPrefixSize}; the two algorithms differ in that the
allocation size inference requires a $<response\ address, target\ address>$
mapping and uses target addresses to infer the LAN network size, while the
rotation pool size calculation requires only the set of response addresses and
uses these to calculate the distance that periphery addresses can travel in an
\ac{AS}.
Rotation
pool size inference can be complicated by devices rotating out of our probe
window, leading to missing observations and erroneously inferring a smaller
rotation pool size than exists in reality.
An \eui \ac{IID}'s rotation
pool is bounded by the entirety of the provider's \vsix allocation (assuming the
\ac{CPE} owner does not change providers), and a /64, meaning that the \ac{CPE}
\ac{IID} was in the same prefix throughout the duration of our study.
This does not preclude a rotation interval longer than our measurement campaign,
however.
\section{Discovering Prefix-Rotating Providers}
\label{sec:campaign}

Our methodology for tracking and traffic correlation 
applies to any \vsix provider, regardless of whether 
they employ prefix rotation (if they do not, tracking is trivial.)
If \eui \acp{IID} in an \ac{AS} to be tracked are
known a priori, the active probing techniques discussed in \S\ref{sec:methodology} 
can be employed to track these devices.
We next consider an 
off-path adversary that wishes to find \emph{all} networks that employ prefix
rotation and track \emph{all} users within those networks.  In
\S\ref{sec:case} we demonstrate the real-world ability of an
off-path attacker (us) to perform this tracking against a set of
clients.

Given the massive \vsix address space, we bootstrap
using existing
large-scale traceroute campaigns to identify networks with \eui addresses at the
network periphery.
We use a traceroute campaign~\cite{caida-routed48} conducted 
from March to April 2019 by CAIDA that issued a traceroute to a single target in each /48
subnetwork of all networks /32 or smaller that were advertised in the
\ac{BGP}.
Even though the CAIDA measurements were taken more than a year before
ours, 
this ``seed'' data
valuably informs our target selection for address discovery and evaluation of prefix
rotation (\S\ref{sec:rotationDetection}) by identifying /48 networks
that produce last responsive hops that have \eui
\acp{IID}.

The CAIDA seed data contains 32,325  
/48 networks with a unique responsive \eui \lhop address -- that is,
no other target address in a different /48 resulted in the same \lhop \eui
address. These 32k /48 networks are part of 938 distinct /32 networks from
627 \acp{AS}; we use
these /32 networks as the starting point of an active measurement campaign to
discover CPE currently using \eui addresses, detect prefix rotation, and ultimately 
track targeted devices.  The subsequent steps included three
tasks:
\begin{enumerate}
 \item Expansion and validation of seed \eui prefixes
 \item Candidate /48 \eui density inference
 \item Prefix rotation detection 
\end{enumerate}

\subsection{Seed /48 Expansion and Validation}
\label{sec:round0}

For the 938 /32 target networks obtained from the seed data, 
we select a random /64 in each of the /32 network's constituent
/48s, and append a random \ac{IID} to produce 
61,472,768 target addresses ($938
\times 65536 \times 1$). As detailed in \S\ref{sec:activeprobing}, we use \zmap to probe 
to each target and
collect
responses. 
This probing step validates that the CAIDA seed data produced
/48 prefixes that generate \eui addresses at the periphery, as well as attempts
to discover other /48s within the same /32 that produce \eui addresses. 
We find 48,970 /48 networks that 
produce a unique \eui
address response (expanding the seed data), and use these in the next
step of probing.

\subsection{Candidate /48 \eui Density Inference}
\label{sec:round1}

We conduct a second round of active measurements to determine the \emph{density} of \eui
periphery addresses within candidate /48 networks. Here, density refers to the
number of unique \eui response addresses received divided by the number of
probes sent to target addresses in the /48.
Different \eui densities can arise from variation in the size of prefix
allocated to the customer. For example, a provider allocating
/48 networks to end sites might have an EUI density of $\frac{1}{2^{64-48}}$,
while on the opposite end of the spectrum, a provider allocating /64s to end
sites may have a theoretical maximum density of $1$ if each /64 is allocated to a
different \ac{CPE} with an \eui address. A prefix's density is influenced both
by the presence of CPE using \eui addresses and the size of the customer
allocations.

We send one probe to each /56 in the /48 network generated from \S\ref{sec:round0}
(approximately 12.5M new probes). 
We aggregate the results by target /48,
counting each unique \eui response generated by probing to a
destination in each target /48. If the number of unique \eui responses 
per target sent (the unique \eui response density) is less than 0.01 (meaning
that the number of unique \eui responses was 2 or fewer),
we classify this target /48 network as low density, and omit it from
future probing. We chose this threshold to eliminate prefixes allocated to a
single device or load balanced between two interfaces; 
while networks that assign
/48s to \ehosts can also rotate prefixes, the exhaustive probing we perform in
\S\ref{sec:rotationDetection} and \S\ref{sec:results} prohibit inclusion of
these networks due to the additional time required to probe them, while yielding
few distinct responses. 
All other networks ($>2$ \eui responses) we classify as high
density, and use them for host discovery probing. 
By this 
conservative
definition, we discover 17,513 high density and 27,429
low density target /48 networks. For the 4,028 remaining networks, we
obtain no responses
after probing them in \S\ref{sec:round0}, and so omit them from
subsequent 
probing.

\subsection{Prefix Rotation Detection}
\label{sec:rotationDetection}

After identifying /48 networks that we are confident
currently generate \eui
responses when probed, we turn to the last step: identifying which networks 
exhibit \emph{prefix
rotation}. We 
concatenate a random \ac{IID} to each 
/64 of the 17.5k high density prefixes
to generate 
$\sim$1.1B target addresses.
We use \zmap to probe each of these addresses in a random order, attempting
to elicit a response from \ac{CPE} routers. We repeat this scan again
in the same order 24 hours later.

With two complete scans of the 1.1B addresses, we filter for $<target,
response>$ pairs where the response is an \eui \vsix address in either scan. Then,
we remove the $<target, response>$ pairs that are common between the two scans. We
do this to isolate targets with an \eui responsive address
that changes over time. These scans reveal \eui address changes, 
potentially indicating a short interval between when a prefix is released by one \ac{CPE}
device and then reallocated to another. Our approach also discovers shifts between a responsive \eui address
and no response, which suggests that the prefix was allocated to a
\ac{CPE} and then returned to the unallocated pool,
or from an \eui address to a
different type of \vsix address, such as \ac{SLAAC} with privacy extensions. We
observed \eui responsive addresses change in 12,885 /48 networks. 

Our approach, where two snapshots are taken 24 hours apart and compared, will
likely fail to capture networks whose prefix rotation interval is on a longer
timescale (\eg 6 months). 
While a provider's rationale for rotating prefixes may vary -- for
instance to prevent static addresses for a particular service tier, to
manage IP reputation, or for privacy -- we focus on networks that 
routinely and
frequently rotate delegated customer prefixes as these are the
most challenging to track from a privacy perspective.
We
also chose not to set a threshold for prefix rotation (\eg 75\% of responsive \eui
addresses leave a prefix between the two snapshots) in order to allow for a
gradual or non-uniform rotation. Such a threshold could be set to further
restrict the definition of prefix rotation, however.  In future work,
we plan to more exhaustively explore the range of provider behaviors,
including rotations on a weekly or monthly basis.

The result of these three steps is a set of $\sim$13k /48 networks that are
likely employing prefix rotation.  
Table~\ref{tab:asn} lists the ASN and countries with the 
most rotating /48 prefixes.
Versatel (AS8881, a large German residential provider) dominates, 
accounting for 40\% of all /48s;
as a consequence, Germany is the most prevalent rotating country
with 46\% of all /48s. AS8881 is a large
residential DSL provider that rotates \ac{CPE} periphery prefixes on a daily
basis, making its prefixes more likely to be flagged as rotators 
than service providers with a longer rotation
period. However, 
more than 100 \acp{AS}, across 25 different countries,
have at least one rotating /48. While this is a small fraction of the total
number of \vsix \acp{AS}, this represents 16\% of the total \acp{AS} with an \eui
address discovered by the original CAIDA routed /48 scan. 

Furthermore, during the three
stages outlined in \S\ref{sec:campaign}, we
discovered 19.4M total \vsix addresses. Of these 19.4M total addresses, 14.8M were
\eui addresses, while 4.6M were not. Only 6.2M \acp{IID} of the 14.8M \eui addresses
were unique, indicating that a significant fraction of the
\eui addresses rotated during this detection phase.

We next
describe a measurement campaign conducted over 1.5 months against these identified
rotating
/48 networks in order
to characterize provider allocation
policies and characterize prefix rotation behavior in the wild.

\begin{table}[t]
 \centering
  \caption{Top 5 ASNs and Countries in Probe Campaign by Number of /48 Prefixes Probed}
 \label{tab:asn}
 \begin{adjustbox}{width=1\columnwidth}
 \scriptsize{
 \begin{tabular}{|c|r||c|r|}\hline
   \textbf{ASN} & \textbf{\# /48} & \textbf{Country} & \textbf{\# /48} \\\hline\hline
   8881 & 5,149 & DE & 5,985 \\ \hline
   6799 & 3,386 & GR & 4,063 \\\hline
   1241 & 635 & CN & 1,126 \\\hline
   9808 & 608  & BR & 561\\\hline
   3320 & 530  & BO & 264\\\hline
   96 Other ASNs & 2,577 & 20 Other Countries & 886 \\\hline
   \textbf{Total} & 12,885 & \textbf{Total} & 12,885\\\hline
 \end{tabular}
 }
 \end{adjustbox}
\end{table}

\section{Measuring Internet-wide Prefix Rotation Behavior}
\label{sec:results}

Having identified 13k /48 networks exhibiting likely prefix rotation
(\S\ref{sec:rotationDetection}), we probed these networks on a
daily basis.
Our measurement campaign spanned 44 days from
late July to early September 2020 and %
probed 844M addresses each day. To ensure temporal consistency
across daily \zmap runs, we probed the same addresses every 24 hours in the same
order (same \zmap random seed), beginning at the same time each day. 

Over the course of our campaign, we sent over 37B \icmpsix Echo Requests to these 844M
destination addresses. We received more than 24B responses from 134M unique \vsix
addresses, including 110M unique \eui addresses. Of these \eui
addresses, we observed only 9M distinct \ac{IID}s, indicating that we observed the
same \eui \ac{IID} 
appearing in many different addresses throughout our
study.\footnote{\S\ref{sec:pathologies} examines pathologies including MAC
reuse}
\subsection{Provider \eui Homogeneity} 
\label{sec:homogeneity}

\begin{figure}[t]
 \centering
  \includegraphics[width=0.9\linewidth]{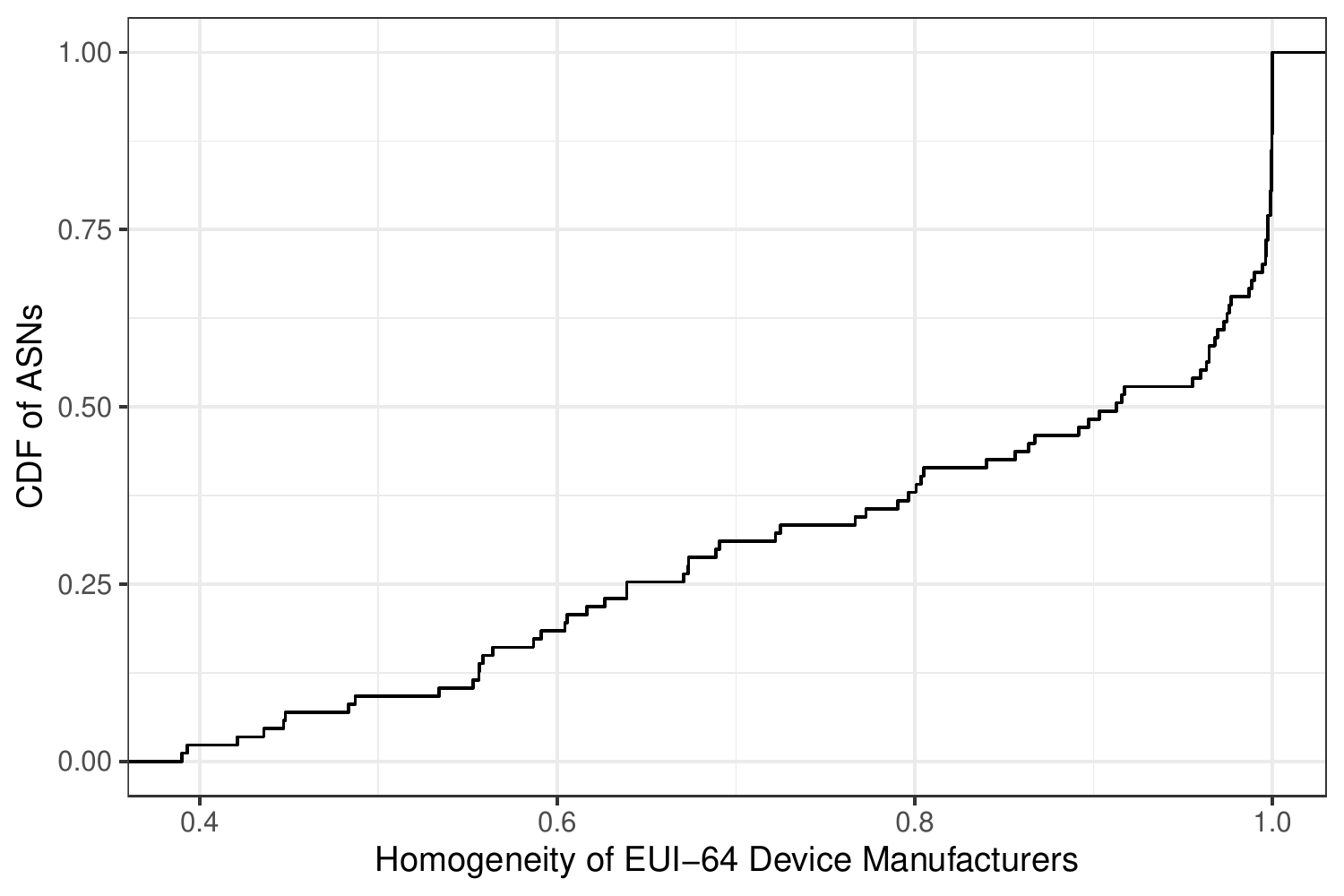}
  \caption{Fraction of devices belonging to the most common manufacturer
  per ASN. Manufacturer homogeneity is indicative of ISP-provided CPE; highly
  homogeneous ASes aid attackers focused on specific device brands or models.
  Eighty-seven ASes are represented; ASes with <100 \eui \ac{IID} are
  excluded.}
  \label{fig:euihomogeneity}
\end{figure}

\eui
\acp{IID} are formed by setting the \ac{U/L} bit of the hardware MAC address,
and inserting \texttt{ff:fe} between the third and fourth bytes.
Hence, it is trivial
to recover the \ac{CPE}'s Internet-facing \ac{MAC} address by reversing
this process. Because the three high-order bytes (commonly known as the
\ac{OUI}) of a \ac{MAC} address encode information about the manufacturer of the
device or interface, we use publicly-available information~\cite{oui}
to study the
per-\ac{AS} distribution of CPE manufacturers we discover.

Service providers exhibit distinct manufacturer fingerprints. For example,
we discovered 205,559 distinct \ac{MAC} addresses embedded in \eui addresses
belonging to NetCologne (\ac{AS}8422), a German ISP. Of these, 99.98\%
(205,527) were in \acp{OUI} belonging to AVM~\cite{avm}, a German
electronics company that produces Fritz!Box \ac{CPE} routers. 
Of the remaining
32 \ac{MAC} addresses, 24 belonged to \ac{OUI} registered to Lancom Systems,
another European \ac{CPE} manufacturer, one was registered to Zyxel
Communications, and the final seven did not resolve to any \ac{OUI} listed by the
IEEE. 

Viettel Group (AS7552), a Vietnamese ISP, similarly exhibited a high degree of
CPE homogeneity, 
but with a different set of
manufacturers. These differences reflect geographic market presence of
device manufacturers, as well as relationships between providers and
the products they lease or sell to subscribers.
We discovered 420,248 distinct
\ac{MAC} addresses within \eui addresses belonging to Viettel prefixes. Here,
99.6\% (418,611) of these \ac{MAC} addresses are registered to ZTE Corporation,
a Chinese manufacturer.

Figure~\ref{fig:euihomogeneity} examines the distribution of 
manufacturers between and within \acp{AS}.
We define an AS's homogeneity as 
the fraction of unique \eui \acp{IID}
belonging to the most common device vendor per \ac{AS}:
$$
\mathit{homogeneity}(ASN) = max\Big(\frac{\text{unique EUI-64 IID(manufacturer)}}{\text{total unique EUI-64 IID}}\Big)
$$
We exclude ASes with fewer than 100 \eui \acp{IID} to prevent \acp{AS} with 
few \eui \acp{IID} from skewing the distribution. 
We observe a that high device
manufacturer homogeneity is common;
in the 87 remaining \acp{AS} these \eui
addresses belong to, more than half have a homogeneity index $>0.9$, and
three-quarters of \acp{AS} have a homogeneity index above $0.67$. 
Even in the least homogeneous \acp{AS}, more than one out of three devices
is manufactured by the same entity. Despite this level of homogeneity within
\acp{AS}, more than 200 distinct manufacturers are observed throughout
the
\acp{AS} we characterize.

Exposing device information via embedded MAC addresses opens up a network
to targeted attacks, for instance for an attacker that 
has identified a vulnerability specific to a certain vendor or
model. Such an attacker might scan networks with many devices
manufactured by that vendor to find potential targets, and
eliminate the need to search networks where few or no devices made by the vulnerable
manufacturer appear.  Further, 
a high-degree of device homogeneity can make individual
network less robust and resilient.

\begin{figure*}[t]
    \begin{subfigure}{.48\textwidth}
     \centering
  \includegraphics[width=\columnwidth]{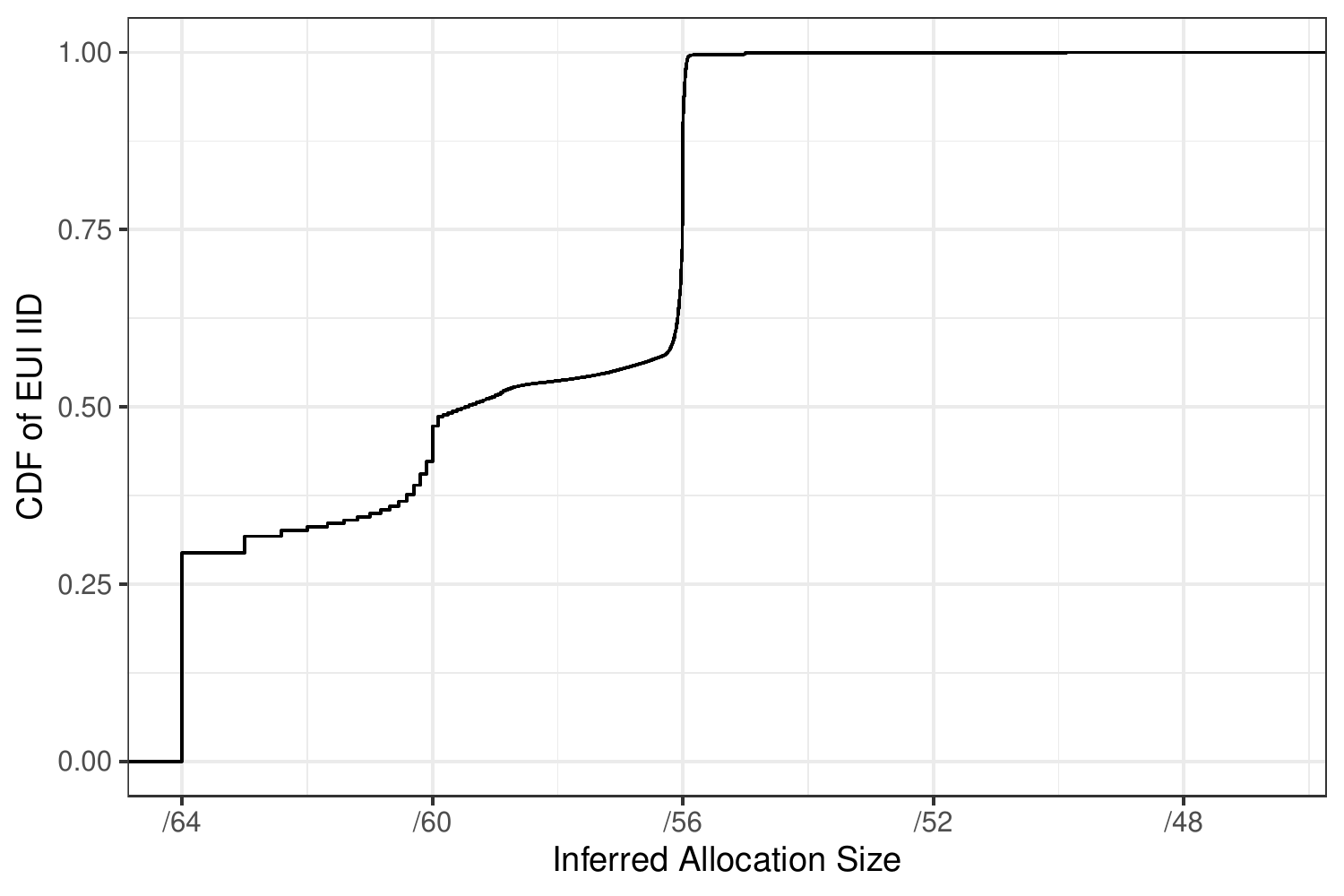}
        \caption{Inferred allocation size of \eui \acp{IID} as a CDF of all \eui
        \acp{IID}. A plurality of \acp{IID} ($\sim$40\%) are allocated /56
        subnets by their service provider, while $\sim$30\% are allocated /64s.
        An inflection point at /60 indicates it is a less-common allocation
        size.}
  \label{fig:allocSizeEUI}
    \end{subfigure}%
    \hspace{1em}
    \begin{subfigure}{.48\textwidth}
     \centering
  \includegraphics[width=\columnwidth]{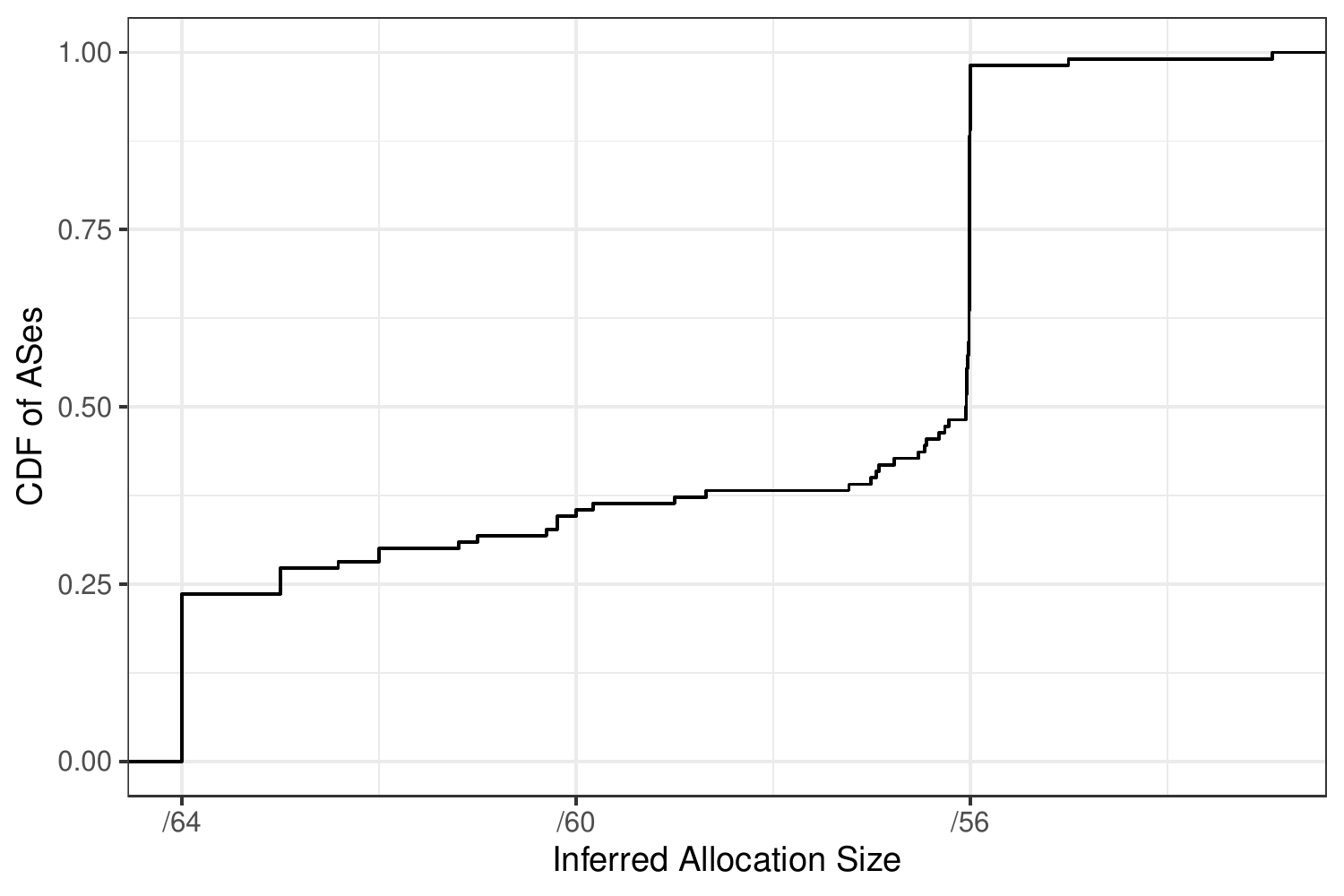}
    \caption{Median inferred prefix allocation sizes of \acp{AS}. A /56 is the most
    commonly allocated prefix size of the providers we probed representing about
    half of the probed ASes. Approximately one-quarter of the ASes
    allocate /64 networks to their customers.}
  \label{fig:allocSizeCDF}
    \end{subfigure}%
    \caption{Inferred Allocation Sizes as CDFs of \eui \acp{IID} and \acp{AS}.}
    \label{fig:allocs}
\end{figure*}

\begin{figure*}[t]
    \begin{subfigure}{.48\textwidth}
     \centering
  \includegraphics[width=.75\columnwidth]{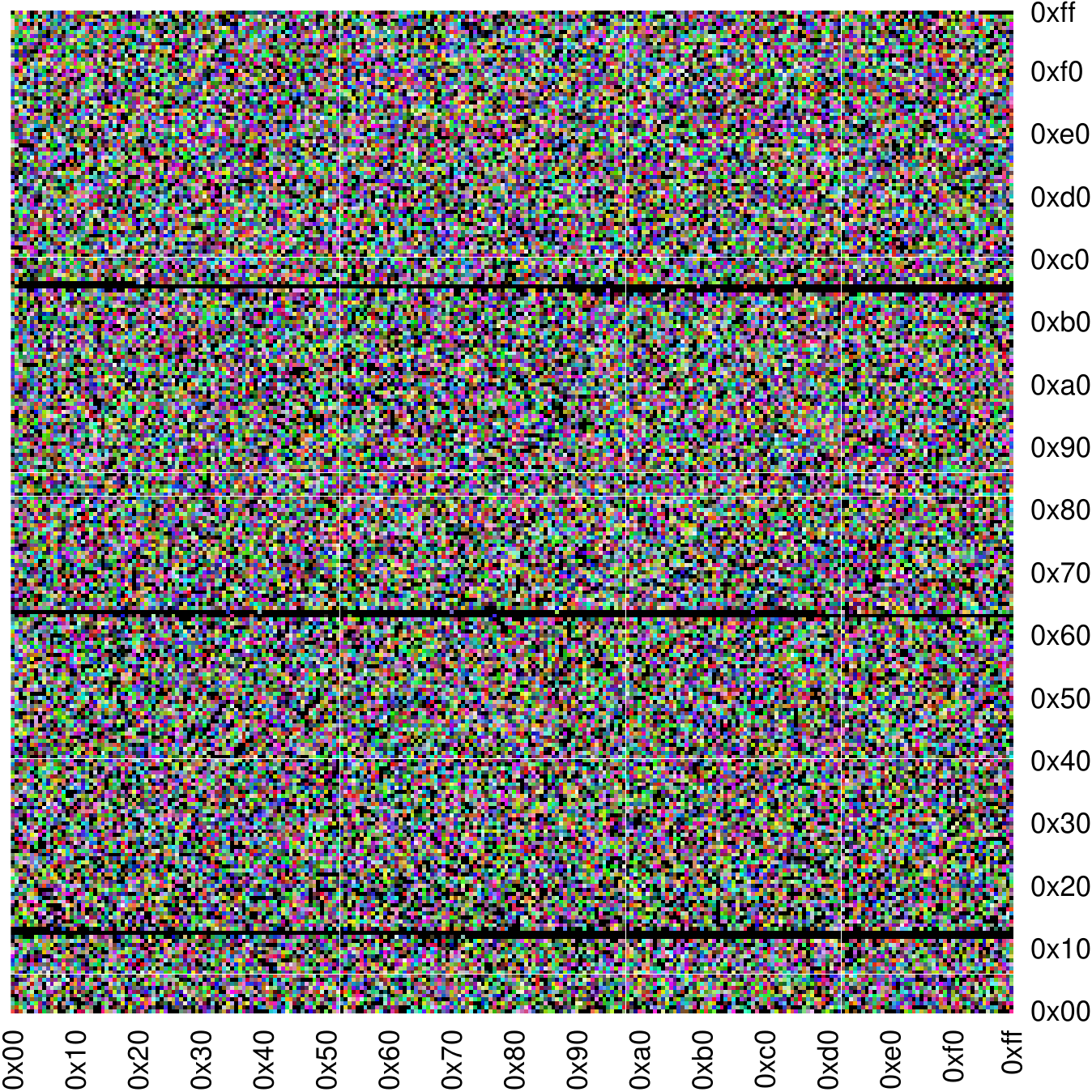}
        \caption{A Versatel prefix (2001:16b8:501::/48) with inferred /64
        customer allocations.}
  \label{fig:multi64}
    \end{subfigure}%
    \hspace{1em}
    \begin{subfigure}{.48\textwidth}
     \centering
  \includegraphics[width=.75\columnwidth]{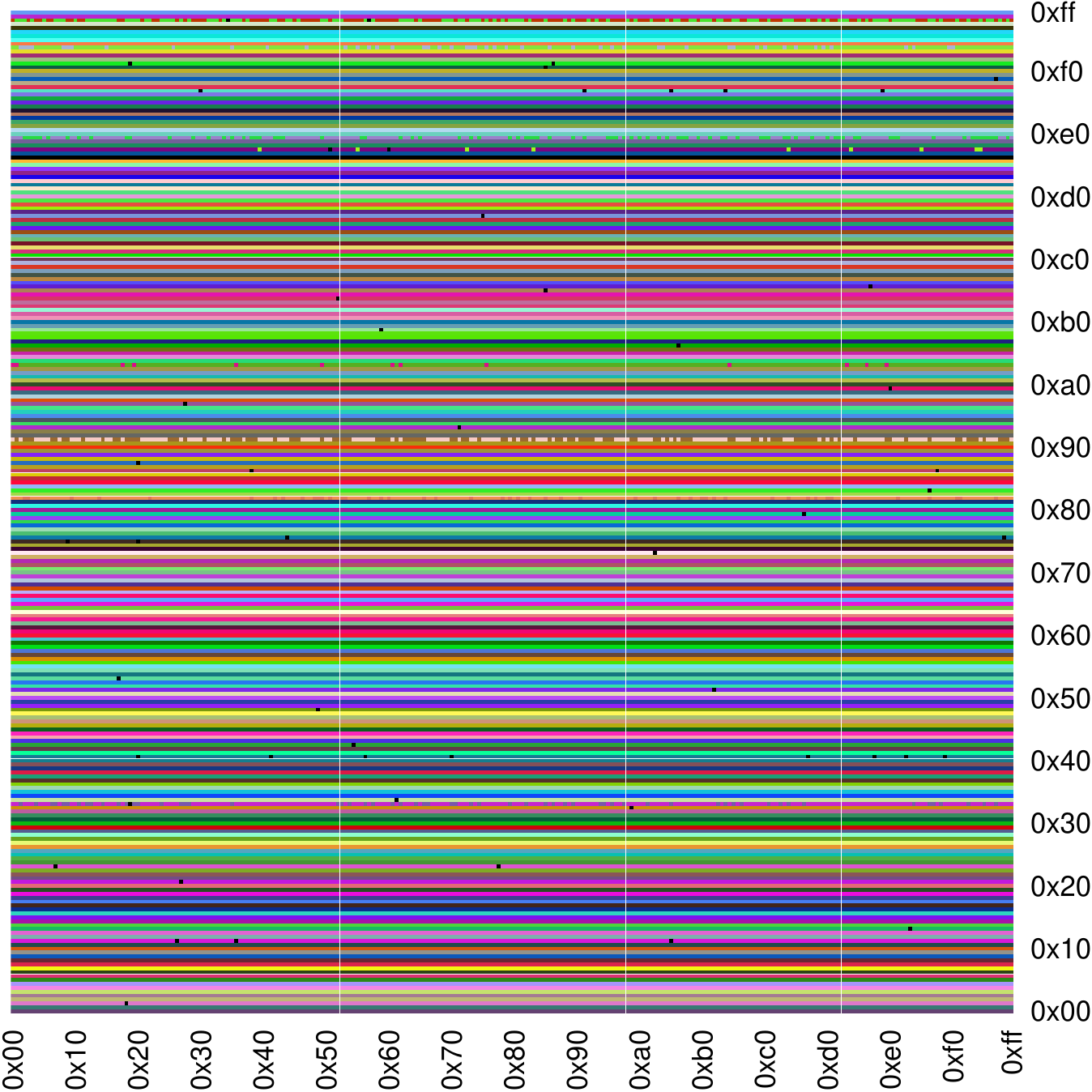}
        \caption{A Versatel prefix (2001:16b8:11f9::/48) with inferred /56
        customer allocations.}
  \label{fig:multi56}
    \end{subfigure}%
    \caption{A provider exhibiting multiple prefix allocation sizes.}
    \label{fig:multisizes}
\end{figure*}

\subsection{Prefix Allocation Sizes}

The subplots of Figure~\ref{fig:allocs} demonstrate our ability to infer the
size of allocations delegated to subscribers by \acp{ISP}. First,
Figure~\ref{fig:allocSizeEUI} displays the CDF of inferred allocation
sizes for all \eui \ac{IID} discovered on a single day of probing. The most
common allocation size delegated to a \ac{CPE} is /56 at about 40\% of all \eui.
The /64 allocation size is also common, at approximately 30\% of all \eui, and
an inflection point at /60 indicates that this size is also allocated, albeit
with less frequency in our data.
Figure~\ref{fig:allocSizeCDF} depicts the cumulative density of
median allocation sizes by \ac{AS}; /56 is the most commonly allocated size
among the providers we probed,
accounting for 50\% of all \acp{AS}.
We infer that about one-quarter of \acp{AS} allocate
a /64 to customers, with another one-quarter using between /64 and /56
prefixes. Some
of these \acp{AS} allocate /60s (\eg BH Telecom, Figure~\ref{fig:sixty}),
others allocate /56 or /64 with noise introduced in the inference by prefix
rotation, and still others provide multiple allocation sizes for different
customers. Figure~\ref{fig:multisizes} depicts two /48s from the same \ac{ISP},
one of which is divided into /56 allocations while the other is split into /64
subnets.

Understanding a provider's prefix allocation size(s) can potentially enable an
adversary to increase their scanning efficiency. For instance, assume that an
attacker attempting to track a \ac{CPE} device that uses \eui addresses knows
that the \ac{AS} the device resides in typically allocates /56 networks to end
sites. The adversary then needs only to probe each /56 subnetwork when
attempting to discover the targeted device. We observed some \acp{AS} 
allocate multiple prefix sizes to end sites, likely due to a variety of customer
types or service plans. In these cases, an adversary may choose to scan
initially assuming the larger allocation size to potentially benefit from scan
efficiency. In the event the targeted device goes undiscovered, a second scan
using the smaller allocation size may be necessary to receive a reply. 

\subsection{Validating Prefix Rotation Detection}
\label{sec:prefixrotation}

\begin{figure}[t]
 \centering
  \includegraphics[width=\linewidth]{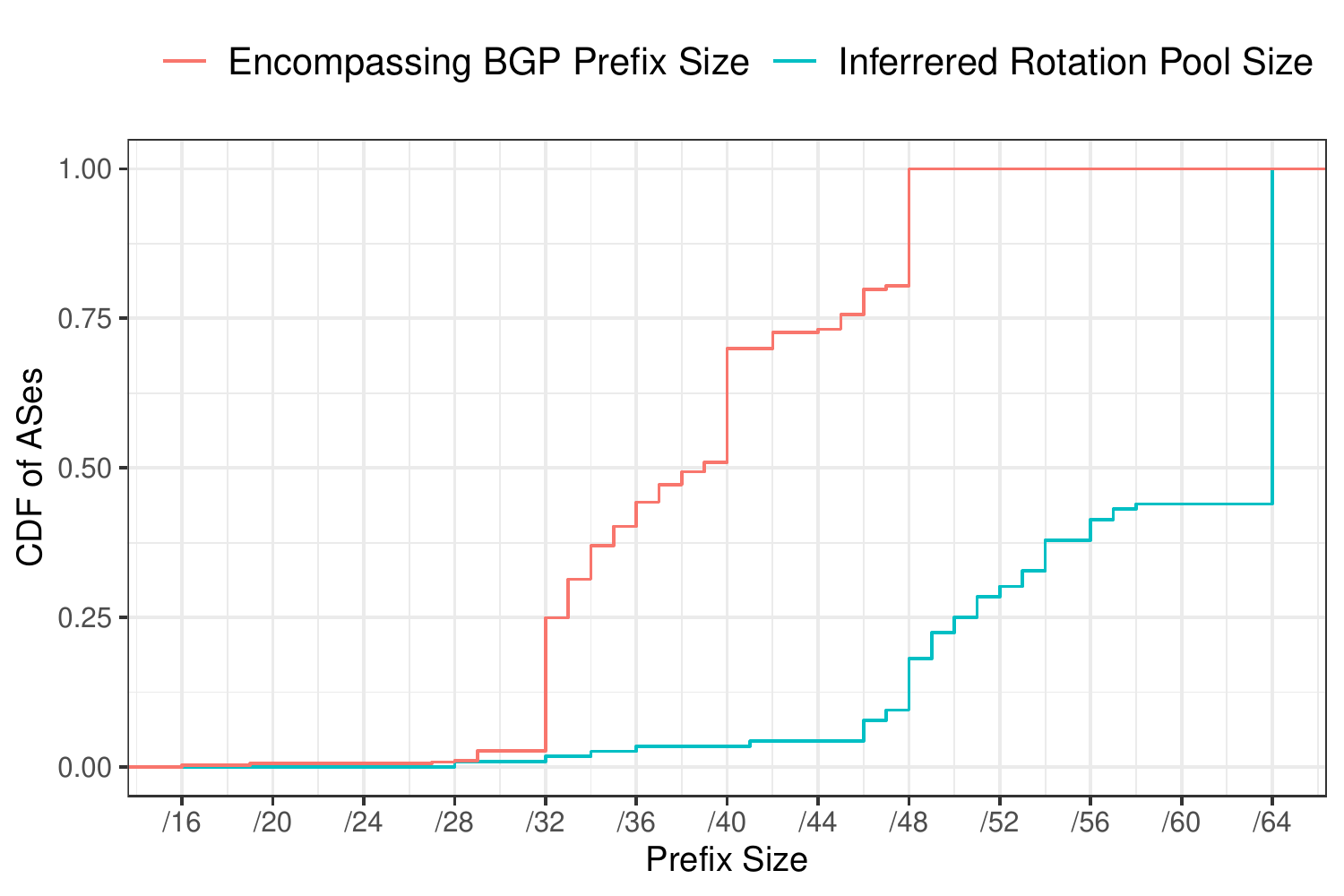}
  \caption{Inferred rotation pool sizes of \acp{AS} vs BGP-advertised prefix
    sizes. The difference between the BGP-advertised prefix and rotation pool
    size represents an attacker's cost savings attempting to track a device with
    an \eui address, as addresses tend to stay within their rotation pools for
    long periods of time.}
  \label{fig:poolSizeCDF}
\end{figure}

Figure~\ref{fig:poolSizeCDF} depicts the inferred rotation pool sizes
for the 101 \acp{AS} to which the \eui addresses belong, as well as the
encompassing BGP prefix sizes for the \eui addresses. We use
Routeviews~\cite{rv} global BGP data to map response addresses to BGP prefixes.

We infer a rotation pool size of /64 for more than half of the
\acp{AS} we probed.
This is indicative of non-rotation within these \acp{AS}, 
/64 is the minimum allocation size possible to allow \vsix hosts to perform
\ac{SLAAC}.
While our initial probing in \S\ref{sec:campaign} was designed to find 
prefixes with rotation pools, it is sensitive to 
the appearance or disappearance of \eui addresses that may be caused
by 
customers joining the prefix or turning off their \ac{CPE}, rather
than larger-scale rotation events. Conversely, approximately half of the
\acp{AS} we probed during our measurement campaign \emph{did} exhibit measurable
prefix rotation, as evidenced by rotation pool sizes $>$ /64. Finally, we compare
the inferred rotation pool size, representing the range within which we \emph{observed} an
\eui \ac{IID} rotate addresses, to the BGP-advertised encompassing
prefixes, which represent the \emph{possible} range within 
which the \eui \ac{IID} might
appear. The difference between the two lines is approximately a /16
throughout, indicating that an \eui \ac{IID} typically rotates within only
$\frac{1}{2^{16}}$ of the possible range. This allows 
an adversary attempting to track an \eui device to bound their search space
to a modest subnetwork of the encompassing BGP prefix, increasing their
likelihood of success.

\begin{figure}[t]
 \centering
  \includegraphics[width=\linewidth]{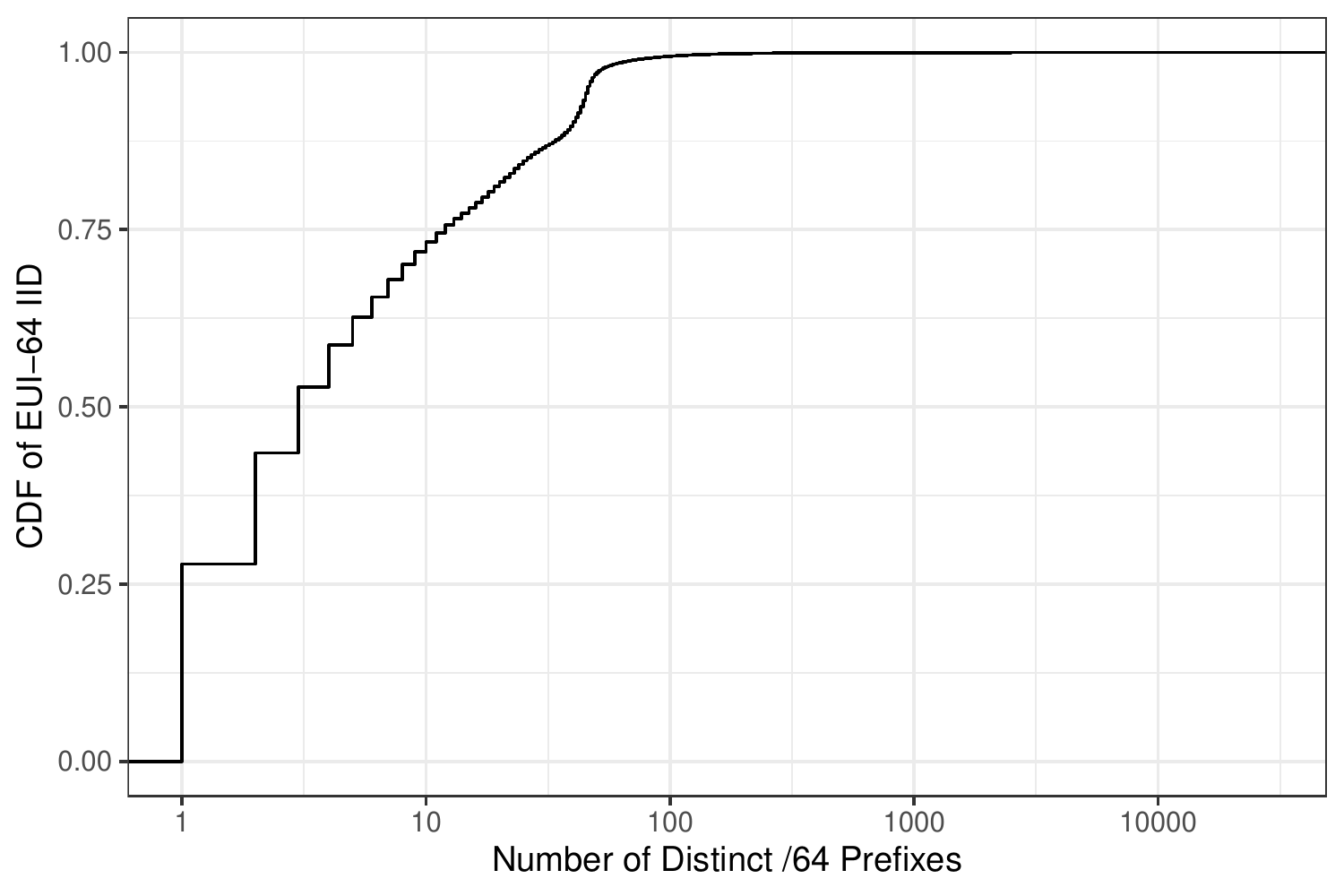}
  \caption{Number of unique /64 prefixes per \eui \acp{IID} ($x$-axis
  log-scale). Most ($\sim$70\%) \eui are found in more than one /64, indicating
  they rotate prefixes during the observation window.}
  \label{fig:prefixesPerIID}
\end{figure}

Figure~\ref{fig:prefixesPerIID} is a CDF of the number of distinct /64 network
prefixes in which we observed each \eui \ac{IID} (alternatively, a distinct \ac{CPE} \ac{MAC}
address) in responses to our probes. We observed about one-quarter of \eui
\ac{IID}s in only one /64 during the course of our study. There are
two potential reasons this might occur. First, the containing address 
never rotates. Although we attempted to identify rotating prefixes
(\S\ref{sec:campaign}), our method may 
erroneously categorize targets as potential rotators due to new
devices appearing between the first and second snapshots in the rotation
identification stage (\S\ref{sec:rotationDetection}). A second explanation 
is that the address \emph{does}
rotate prefixes, but the prefix that it rotates to does not belong to the set of prefixes
that we probe daily. In this case, we would not observe the \ac{IID} for some
period of the measurement campaign. More than 70\% of distinct \ac{IID}s 
appeared in more than one /64 prefix during our study, indicating that
they rotated prefixes at least once. Finally, a small number of \acp{IID}
rotated across an extreme number of /64 prefixes,
including one \ac{IID} that appeared in nearly 30,000 distinct /64 networks. We
discuss these anomalies in greater detail in \S\ref{sec:pathologies}.

\subsection{Rotation Pool Behavior}

\begin{figure}[t]
 \centering
  \includegraphics[width=\linewidth]{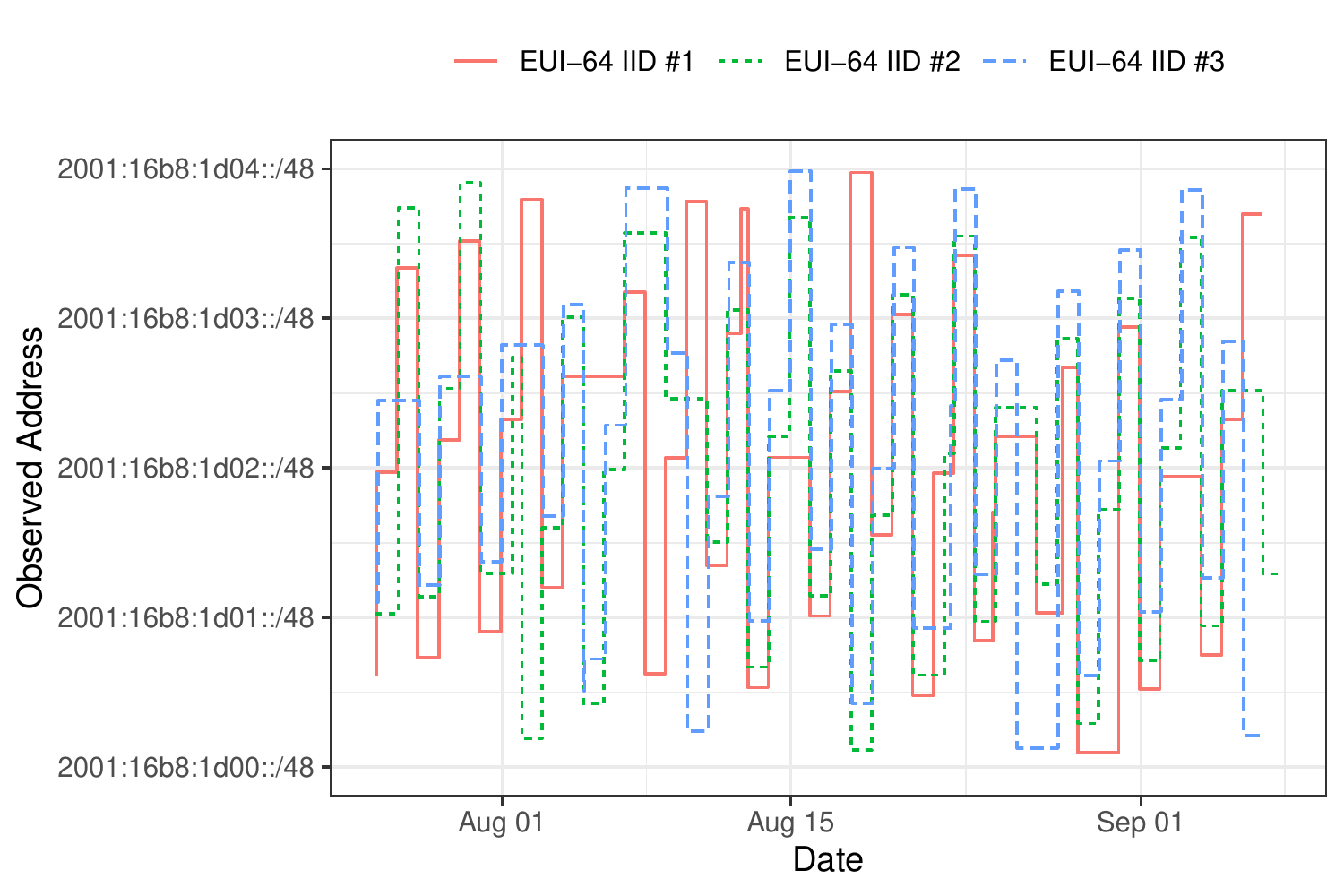}
  \caption{Three AS8881 \eui \acp{IID}' assigned /64 prefixes over time. Each
  \ac{IID}'s network prefix increments each day modulo the size of the rotation
    pool (/46).}
  \label{fig:versatelMACRotation}
\end{figure}

To better understand the lower-level dynamics exhibited by
prefix-rotating providers, we investigated the behavior of \eui addresses
\emph{within} rotation pools. Figure~\ref{fig:versatelMACRotation} displays the
observed /64 prefixes for three \eui \acp{IID} in AS8881.
Each \ac{IID} rotates
throughout the observation window within the same /46, AS8881's inferred rotation pool
size (\S\ref{sec:rotationpool}). The figure shows that each \eui \ac{IID}'s
/64 prefix increments each day; when the value is greater than the /46
rotation pool, the /64 prefix simply wraps modulo $2^{18}$ to remain within the
/46. This causes \eui \acp{IID} \#1 and 2 (red and green lines) to appear in three
/48 prefixes before wrapping modulo the /46 rotation pool for the first several
days, while \eui \ac{IID} \#3 alternates between 2001:16b8:1d01::/48
and 2001:16b8:1d02::/48. This helps scope an
attacker's prediction of what prefix an \ac{IID} will have in the future.

\begin{figure}[t]
 \centering
  \includegraphics[width=\linewidth]{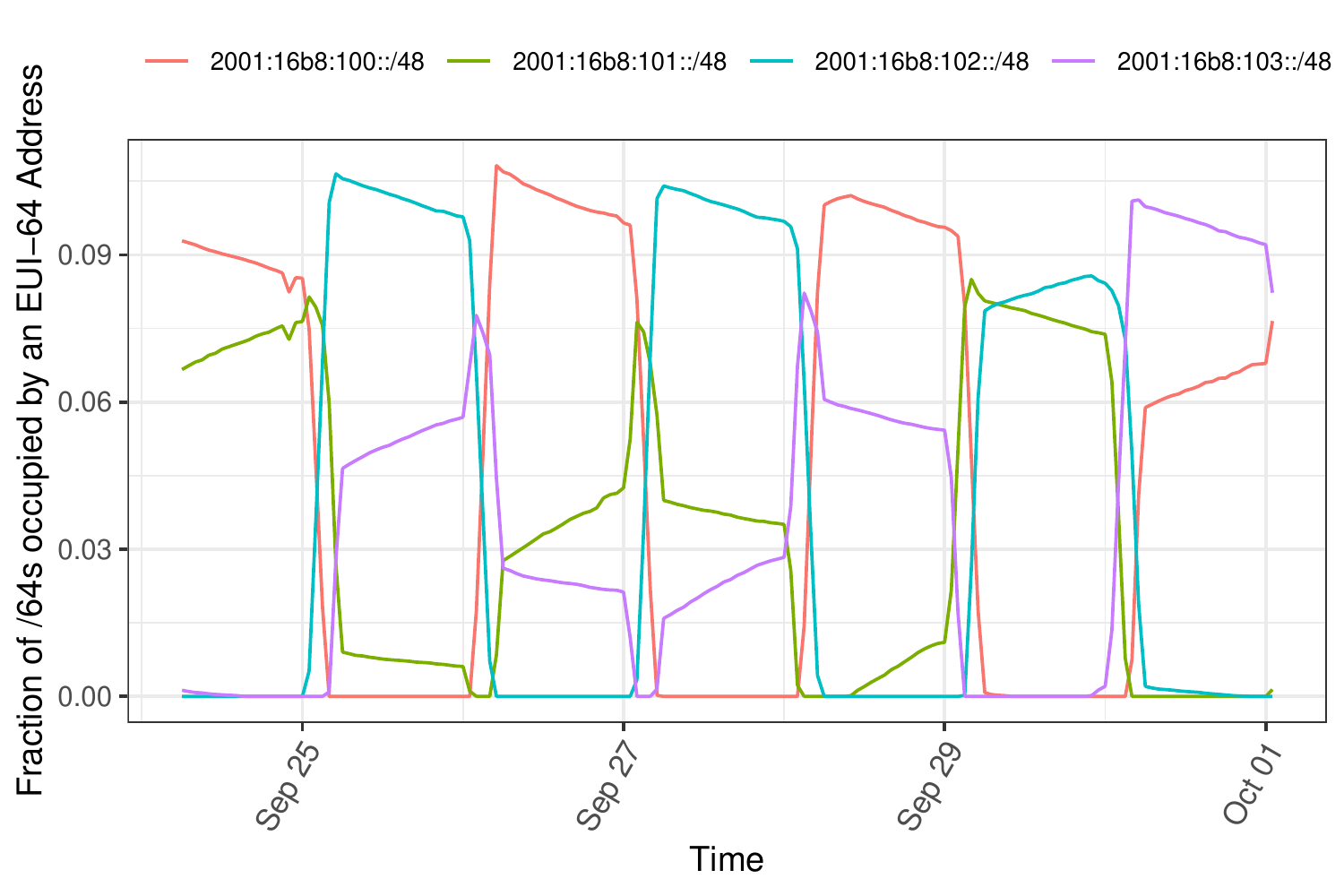}
    \caption{Address density of an AS8881 (DE) /46 rotation pool over time.
    Prefix reassignment typically occurs during the early morning hours; on any
    given day, one prefix contains the majority of \eui addresses between them,
    another contains almost none, and the remaining two are changing densities
    in opposing directions.}
  \label{fig:versatelbyday}
\end{figure}

Next, we look at
the number of \eui addresses in each /48 within a rotation pool, again focusing
on AS8881. 
Figure~\ref{fig:versatelbyday} plots the density of \eui addresses within
each /48 of a /46 rotation pool. For each hour over the course of a week, we
probed portions of the AS8881 address space to elicit responses from \eui
addresses within the 2001:16b8:100::/46 prefix.  Figure~\ref{fig:versatelbyday}
shows that prefix reassignment predominantly occurred between 
midnight and 0600 CEST each day. 
Generally, on a given day, one /48 prefix
contained the most addresses, one /48 contained close to none, and
the other two /48s contained the smaller remainder, changing densities in
opposite directions from each other.

\subsection{Pathologies}
\label{sec:pathologies}

We observed several phenomena during our campaign that bear further
study, as they fell outside of the behaviors we expected \eui \acp{IID} to exhibit.
First, of the 9M distinct \eui \acp{IID} we observed, 10k of these were observed in
multiple \acp{AS}. One, corresponding to the \ac{MAC} address
\texttt{00:00:00:00:00:00}, was observed in 12 distinct \acp{AS}, likely because
this \ac{MAC} is used as a default address, particularly 
when the interface may not have a pre-programmed \ac{MAC} address (such as a
cellular interface on a hotspot device.) 

\begin{figure}[t]
 \centering
  \includegraphics[width=\linewidth]{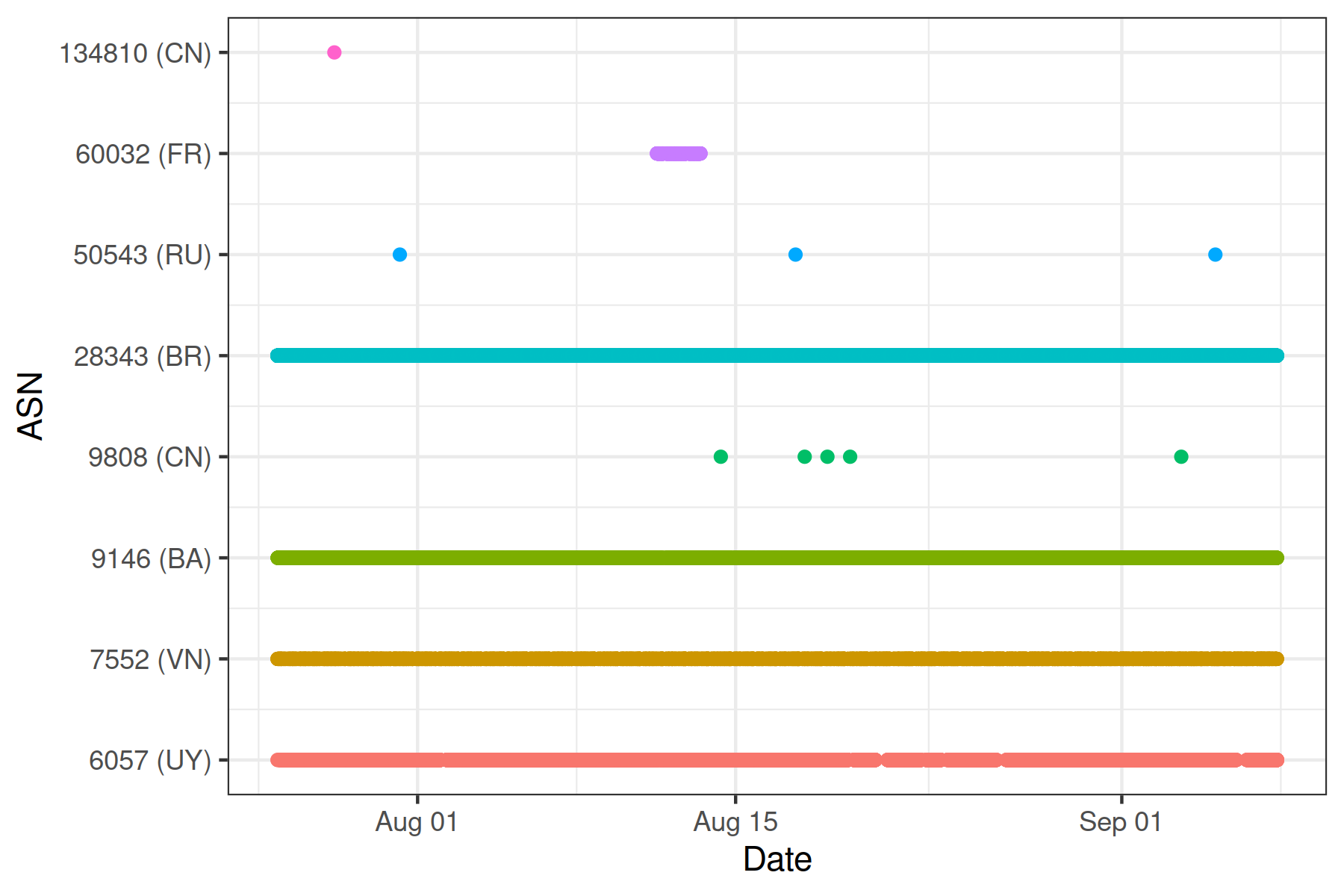}
  \caption{Responses from a single \eui \ac{IID} by ASN over time. Sustained
  observations in multiple countries throughout the observation window suggests
  \ac{MAC} address reuse by a vendor, decreasing its utility as a
  trackable identifier.}
  \label{fig:multiASN}
\end{figure}

\begin{figure}[t]
 \centering
  \includegraphics[width=\linewidth]{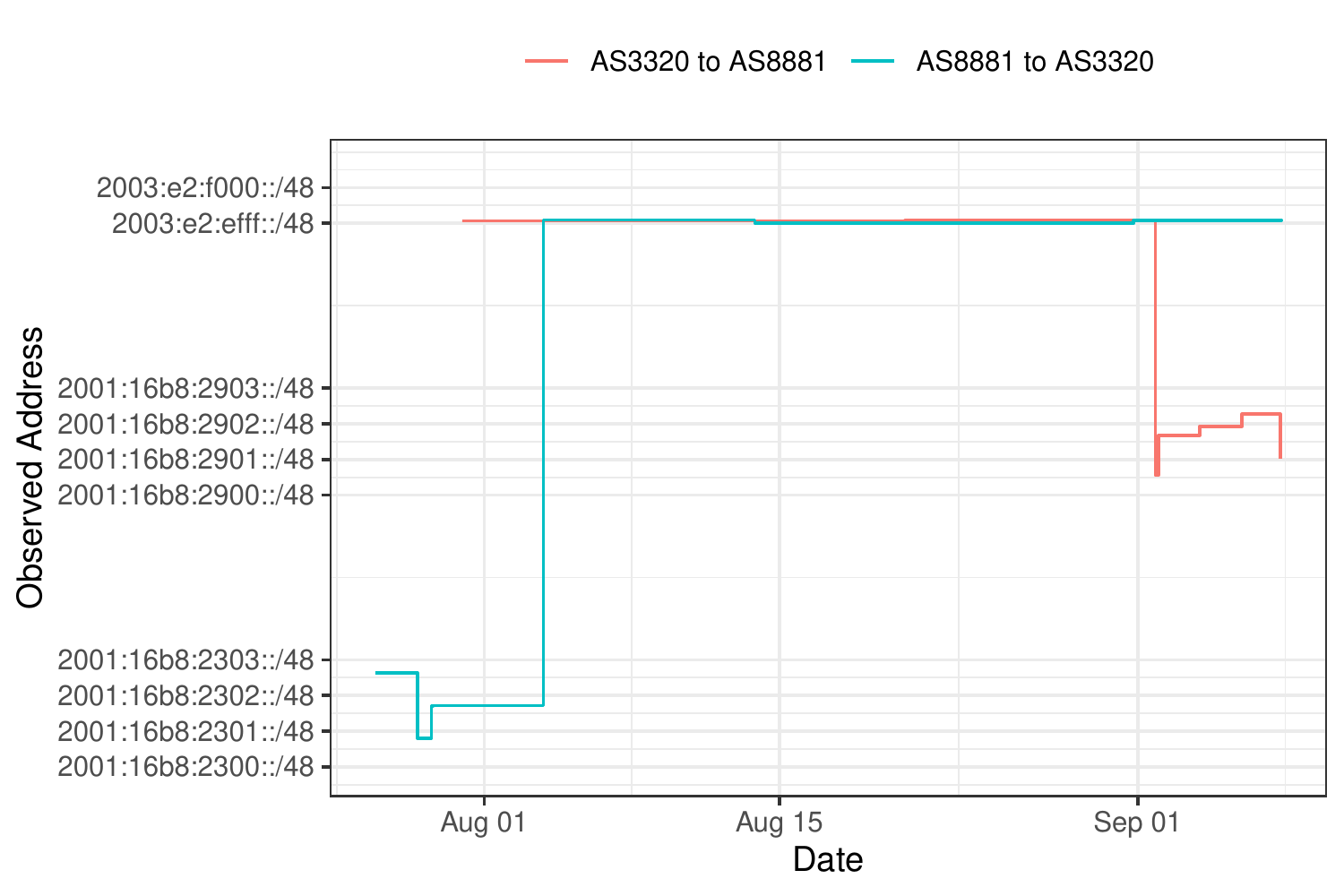}
  \caption{Two \eui \ac{IID} changing between German \acp{ISP} ($y$-axis not to
    scale). Neither are seen in their former provider's network after shifting
    to their new network, indicating that their owners changed service providers
    rather than are utilizing a backup link during an outage.}
  \label{fig:providerChanger}
\end{figure}

Another category of multiple-\ac{AS}
\eui \acp{IID} is more difficult to explain;
in this subset, the same \acp{IID} appeared in networks on different continents within the same day for the
duration of our study. Most of these \eui \acp{IID} correspond to a
single manufacturer. 
Figure~\ref{fig:multiASN} depicts the observations of a single \eui address in
multiple \acp{AS} distributed globally during our measurement campaign. 
In this case, we observed an \eui \ac{IID} appear daily in \acp{AS} located in
Uruguay, Vietnam, Bosnia, and Brazil, as well as occasionally in China,
Russia, and France. Due to the geographic distribution of the \acp{AS}, this is
likely a \ac{MAC} address reused multiple times by the
manufacturer in violation of the standard~\cite{ieee802}. Finally, many of these 10k addresses that appear in multiple
\acp{AS} shift from one \ac{AS} to another, i.e., stop appearing
in the first \ac{AS} after the switch. This behavior is evidence that the
customer has changed service providers, especially if the providers
serve the same region. Figure~\ref{fig:providerChanger} illustrates this
behavior as seen from two different devices, one of the two moving from AS8881
(2001:16b8::/32 addresses) to
AS3320 (2003:e2::/32 addresses) in early August, and the other in the opposite direction in early
September.  Neither are seen in the previous \ac{AS} again after being observed
in the new \ac{AS}. Because both \acp{AS} are German residential service
providers, this behavior appears to represent either a customer switching Internet service
plans, or an instance in which the backup provider became the primary for a
dual-homed device.

\section{Device Tracking Case Study}
\label{sec:case}

\begin{figure*}[t]
  \begin{subfigure}{.48\textwidth}
    \centering
    \includegraphics[width=\columnwidth]{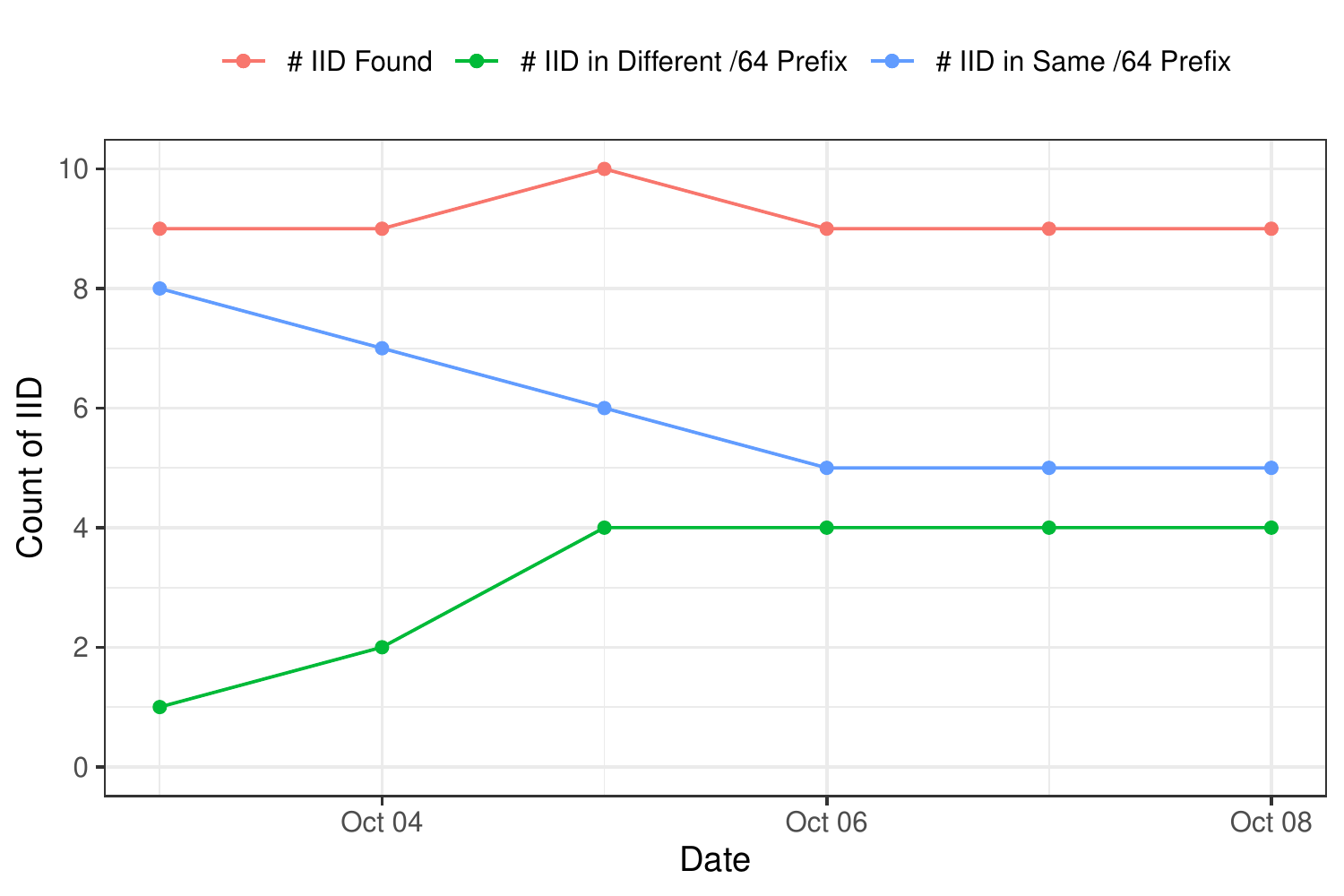}
    \caption{Discovered \eui \acp{IID} during targeted tracking. While some
    \ac{IID} rotate prefixes, others do not; regardless of rotation, 9 of 10 are
    found consistently over the course of a week.}
    \label{fig:trackingCounts}
  \end{subfigure}%
    \hspace{1em}
  \begin{subfigure}{.48\textwidth}
    \centering
    \includegraphics[width=\columnwidth]{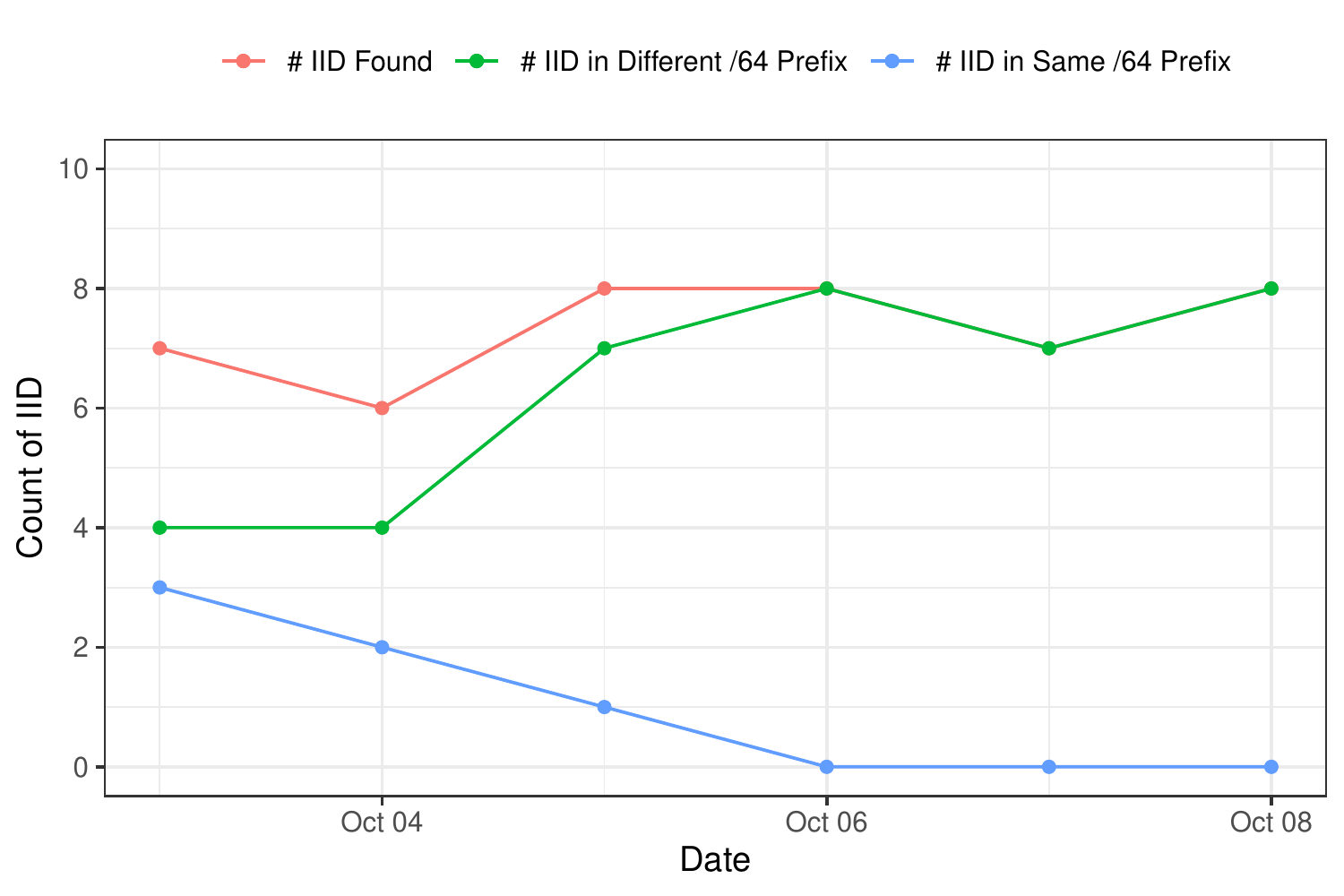}
    \caption{Discovered prefix-rotating \eui \acp{IID} during targeted
    tracking. Even when targeting only devices that rotate prefixes, we
    can track most \acp{IID} during the week-long observation window.} 
    \label{fig:rotationTrackingCounts}
  \end{subfigure}%
  \caption{Results from tracking two sets of ten \eui \ac{IID} over a week.}
\end{figure*}

\begin{table*}[t]
  \caption{Characteristics of ten prefix-changing \eui \acp{IID} tracked over one
  week}
  \label{table:10rotatingstats}
  \centering
  \scriptsize{
\begin{tabular}{|l|l|l|l|l|l|l||l|l|l|l|l|l|l|l|l|}
\hline
\textbf{\eui IID} &
  \textbf{\begin{tabular}[c]{@{}l@{}}Mean Probes / \\ StdDev\end{tabular}} &
  \textbf{\begin{tabular}[c]{@{}l@{}}BGP \\ Prefix\end{tabular}} &
  \textbf{ASN} &
  \textbf{CC} &
  \textbf{\# Days} &
  \textbf{\begin{tabular}[c]{@{}l@{}}\# /64 \\ Prefixes\end{tabular}} &
  \textbf{\eui IID} &
  \textbf{\begin{tabular}[c]{@{}l@{}}Mean Probes / \\ StdDev\end{tabular}} &
  \textbf{\begin{tabular}[c]{@{}l@{}}BGP \\ Prefix\end{tabular}} &
  \textbf{ASN} &
  \textbf{CC} &
  \textbf{\# Days} &
  \textbf{\begin{tabular}[c]{@{}l@{}}\# /64 \\ Prefixes\end{tabular}} \\ \hline
\textbf{\#1} & 4238 / 0             & /48 & 7552   & VN & 7 & 2 & \textbf{\#6}  & 28,331.5 / 10,867.3 & /32 & 56044  & CN & 2 & 2 \\ \hline
\textbf{\#2} & 125,679.7 / 95,694.0 & /32 & 8422   & DE & 7 & 7 & \textbf{\#7}  & 9,293 / 5,575.6     & /32 & 9146   & BA & 7 & 6 \\ \hline
\textbf{\#3} & 379 / 315.7          & /32 & 262557 & BR & 7 & 2 & \textbf{\#8}  & 14,523 / 4,558.5    & /40 & 8881   & DE & 2 & 2 \\ \hline
\textbf{\#4} & 41,413.3 / 17,522.0  & /37 & 27699  & BR & 7 & 3 & \textbf{\#9}  & 1,363.3 / 530.3     & /32 & 10834  & AR & 5 & 2 \\ \hline
\textbf{\#5} & 152,676 / 82,977.9   & /33 & 14868  & BR & 7 & 2 & \textbf{\#10} & 860.3 / 284.8       & /48 & 200924 & DE & 3 & 3 \\ \hline

\end{tabular}
}
\end{table*}

In order to demonstrate our technique's efficacy for tracking \vsix \ac{CPE}, we
conduct an experiment that emulates an adversary interested in tracking specific
\eui \acp{IID} as they change prefixes.  
By utilizing the previously described techniques, this section
demonstrates the real-world feasibility of an attacker re-identifying
traffic from clients despite the deployment of technologies intended to prohibit
such re-identification.

Over the course of a day, we probed 
the targets described in
\S\ref{sec:rotationDetection} to discover current \eui addresses. Then, we
selected ten \eui \vsix addresses at random, with the caveats that
no two addresses came from the same
country or \ac{AS}, and excluding \eui \acp{IID} that we had observed previously in
multiple \acp{AS}
(\S\ref{sec:pathologies}). We then probed over an additional
six days to attempt to track these \acp{IID} for one week.

In addition to demonstrating that \ac{CPE} can be tracked over time, another
central aim of this case study is to validate the space-reduction approach
depicted in Figure~\ref{fig:space}. For each
of the ten selected addresses, we leveraged the allocated prefix size and
rotation pool size inferences we made for each \ac{AS} in \S\ref{sec:allocsize}.
We used the allocation size inference in order to reduce the number of probes a
na\"ive search algorithm might send: one probe per /64 in the suspected target
prefixes. As an example, for providers who commonly allocate /56 prefixes to
customers, we need send only one probe to an address within that subnet in order
to elicit a \ac{CPE} reply -- thereby sending only $\frac{1}{256}$ the number of
probes of a na\"ive scanner.
We incorporated the inferred rotation pool sizes to ensure
we probe the space an address is \emph{likely} to rotate to. 
For example, some \acp{AS} use /46 rotation pools and 
we chose a target in each allocation size
block throughout the entire pool. While this somewhat offsets the cost savings
we obtain by probing only once per allocation
size, it widens the target aperture to encompass the range of addresses 
within which an \eui
address is likely to fall. This approach increases the number of observations
and decreases the likelihood that we ``lose track'' of an \eui \ac{IID}.

Figure~\ref{fig:trackingCounts} displays the number of \acp{IID} 
successfully tracked each day.
Over all six days, 
we discovered 9 to 10 \ac{IID}
(out of 10) targets.
This suggests that, across device manufacturers, 
countries, and \acp{AS}, tracking  \ac{CPE} over time via \eui
addressing is quite feasible.

We also observed that the number of \acp{IID} whose
prefixes changed during the course of the study increased from 1
to 4 \acp{IID}. This matches our
intuition, in that devices may only rotate prefixes after several
days.  
Half of the \acp{IID}
we selected remained in the same prefix over the duration of the
experiment. This
likely indicates that these addresses were not in networks that rotate
prefixes sub-weekly and whose addresses are therefore trivially tracked.

Rather than selecting \eui addresses at random, 
we next chose 10 \eui \acp{IID} that \emph{did}
exhibit prefix rotation during our tracking time frame.
Figure~\ref{fig:rotationTrackingCounts} plots the number of \acp{IID} we
discovered during the study; in this case, we discovered slightly fewer \acp{IID} per
day than without the rotation restriction, with a maximum of 8, and a minimum
daily \ac{IID} discovery of 6 of the 10 \acp{IID}. Unlike
Figure~\ref{fig:trackingCounts}, Figure~\ref{fig:rotationTrackingCounts}
shows that all of the \acp{IID} we tracked eventually changed prefixes by the fourth
day of probing. Despite this, we still discovered 8 of the 10 \eui \acp{IID},
emphasizing that \eui addressing undermines the privacy goals of prefix
rotation.

Table~\ref{table:10rotatingstats} summarizes the characteristics of the 10
prefix-rotating \eui \ac{IID}s tracked over a week, including the 
number of probes sent until they were located in their new prefix.
If the \eui \ac{IID} was not found on a given day,
we include the total number of probes that we sent.
As an example, we discovered ``\eui IID \#3'' after sending only 379 probes on
average each day. Because ``\eui IID \#3'' is part of a BGP-advertised /32
allocation, exhaustively searching the space would entail up to $2^{32}$ probes in
order to enumerate each /64 that could potentially be allocated to a \ac{CPE}
device. At a rate of 10k packets per second, this probing would take nearly five
days, during which period the \eui IID might again change prefixes. By contrast,
using our target selection method and inferred prefix allocation and rotation
pool sizes, we discover this \eui \ac{IID} over the course of a week using only
2,274 total probes in two different \eui \vsix addresses. ``\eui IID \#2'', which
was discovered in 7 different /64 networks over the week of probing, took only
754k packets total to locate, equivalent to 75 seconds of active probing at
10k packets per second.

While the majority of the \eui \ac{IID}s were discovered each day of the tracking case
study, some were not, and two were observed on only two days during the week. 
We consider two explanations for this behavior. First, our inferences of the
provider's allocated prefix and rotation pool sizes may be incorrect. If one (or
both) of these inferences is incorrect, our target generation technique, which
sends one probe to each allocation-sized block within the prefix rotation pool,
may not probe to an address within the new allocation assigned to the \ac{CPE}.
For instance, if we infer a /56 allocated prefix size for a provider, but the
\ac{CPE} is actually within a block that receives a /64 for its internal
subnetwork, it is unlikely that our probes will target the \ac{CPE}
/64 as the probed /56 is chosen at random. Alternatively, if we underestimate
the prefix rotation pool size, the \ac{CPE} may rotate out of the address
space we are probing. If the address pool is 
larger than our estimate, then in the future, we may observe this \eui again
when it rotates back into our probing range. If, however, a device
\emph{changes} rotation pools, the \eui will go unobserved for the duration of
the probing unless some other event forces it back into the probed rotation
pool. In this case, a motivated adversary may attempt to simply identify many or
all of a provider's rotation pools, and enumerate those to find the device
again. 

A second explanation for the disappearance of an \eui \ac{IID} is the removal of
a device from service, whether by a change in provider
(\S\ref{sec:pathologies}), or an extended outage.
In these instances, an adversary would not expect to see the device
return, and should cease probing this \eui \ac{IID} after some time without positive
results.
\section{Ethical Considerations}
\label{sec:ethics}

We followed established ethical principles and recommended practices for high-speed
Internet probing during this work~\cite{durumeric2013zmap}. Our probing
consisted of sending \icmpsix Echo Requests using zmap6. \icmpsix messages are
used for diagnostic and error-reporting purposes, and are considered less
obtrusive than UDP or TCP probes. In order to 
minimize the risk of \icmpsix rate
limiting, which is mandatory, and to 
reduce the load on transit and destination
networks, zmap6 randomizes the order of probing. We probed at a
conservative rate, coordinated with the administrators of the vantage point, and
ran an informative website from the vantage point providing 
contact information to opt-out of our experiment. We received no opt-out requests.

We contacted and initiated a security report with the \ac{CPE} manufacturer
whose devices comprise 22\% of the \eui \ac{IID} we discovered to disclose the 
privacy
vulnerability. We sought to work with device manufacturers to bring awareness to
the tracking implications of \eui \ac{CPE} addresses and to remediate the
privacy flaw by ensuring future iterations of their \ac{CPE} OSes implement
\ac{SLAAC} privacy extensions (\S\ref{sec:remediation}). 

\section{Remediation}
\label{sec:remediation}

\eui \ac{SLAAC} addressing in \ac{CPE} devices enabled the targeted tracking
case study of \S\ref{sec:case}.  The most straightforward way to
protect consumers from this type of tracking is for \ac{CPE} manufacturers to
ensure their products are capable of \ac{SLAAC} privacy extensions and to enable
privacy extensions by default. Because service providers often partner
with \ac{CPE} vendors to offer their products directly to customers (as evidenced
in \S\ref{sec:homogeneity}), they also have an active role to play in
ensuring these devices provide an adequate level of privacy. 

We contacted a \ac{CPE} manufacturer that appeared prominently ($\sim$2 million
MAC addresses) in our
results regarding the privacy and tracking
implications of their continued use of \eui \ac{SLAAC} addresses. A
company representative
informed us that the motivation for their use of \eui \ac{SLAAC} addresses
is to optimize connection setup time.   Specifically,
implementing \ac{SLAAC} with privacy extensions requires the device to
perform \ac{DAD}~\cite{rfc4862,rfc4941}. We believe that the benefit of using
the privacy extensions of \ac{SLAAC} addresses outweighs the delay incurred while
\ac{DAD} is performed, particularly in light of RFC 4429 \cite{rfc4429}, which allows an
interface to be used while \ac{DAD} completes. When presented with our results,
the manufacturer agreed that the privacy vulnerabilities of continued \eui use
outweighed the marginally-increased interface setup time, and indicated that
they
will implement \ac{SLAAC} with privacy extensions rather than \eui \ac{SLAAC} in the next release of their
\ac{OS} in early 2022~\cite{anonymousemail}.

Further, RFC 4941 \cite{rfc4941} specifies only that a device implementing \ac{SLAAC}
with privacy extensions \texttt{SHOULD} generate a new, random \ac{IID} each
time its network changes. Our work shows that \texttt{SHOULD} is too weak, and
the privacy goals of this standard dictate that the 
\ac{CPE} \texttt{MUST} do so in order to prevent the same type of tracking using
the randomized \ac{IID} rather than an \eui \ac{IID}.

In the event that a device cannot support \ac{SLAAC} privacy extensions due to
some technical limitation (unlikely, given that \vsix privacy extensions were
introduced in 2001), ISPs should provide a mechanism to inform users about their
increased vulnerability to tracking. 

\section{Summary and Conclusions}
\label{sec:conclusions}

We used large-scale active measurements to demonstrate a widespread 
vulnerability with deployed IPv6 privacy enhancing technologies
that randomize \acp{IID} and rotate address assignments 
assigned to a single customer.   
Although these enhancements are ubiquitously deployed at 
the edge, the CPE market lags behind. The legacy IPv6 standard 
does not require the use of \ac{SLAAC} privacy extensions, 
so these CPE devices are still compliant with the IPv6 standard; 
however, \eui addressing undercuts modern IPv6
privacy features supported by all major modern \acp{OS}.  As a result, any adversary
that can send active probes toward
random IP addresses in target prefixes, is potentially able 
to track devices in that prefix. 
An adversary could use similar scalable active probing 
methods to identify which IPv6 prefixes are likely to be fruitful targets 
for such tracking.  

We demonstrated methods for both approaches and
applied them to a case study of device tracking, finding the problem spans
CPE devices observed in over 100 ASes and 25 countries.
Simply put, our measurements show that the privacy/anti-tracking mechanisms used in IPv6 today
do not work without additional coordination with the CPE.
Based on our findings, a major CPE vendor, whose products account for over 2
million distinct MAC addresses in our corpus, has deprecated \eui addressing in
its forthcoming \ac{OS} release. This change will materially improve user privacy.

The security and privacy community has devoted extraordinary effort over
the last decade to developing privacy-enhancing technologies intended for
pervasive and general-purpose use.  To be effective, such efforts must
often span technical, policy, and standards development work.  But without
empirical study of how these technologies are deployed,
we have limited understanding of which privacy-enhancing
technologies are truly achieving their intended goals. 
This study represents an example of the interplay between 
privacy-enhancing technologies, and the context in which 
these technologies are deployed, which in many cases can 
inadvertently subvert the privacy enhancements.  Worse, in the
example we studied, users receive
no notification of this interaction, which could cause harm
relative to not having the privacy enhancements in the first place,
because the enhancements offer a false sense of privacy and security.
We believe the research and standards development community should consider
deprecating legacy \vsix behavior that uses static \eui addresses, 
or at least requiring notification to the user of privacy implications.

Content and service providers must also account for
prefix rotation when blocking attack traffic originating from \vsix users. The
\vfour paradigm of denying or rate-limiting a single address or range of
addresses is ineffective when client prefixes may rotate daily.
Our results suggest that future work is needed to better understand how providers 
might rethink employing such security mechanisms in the face of these
deployed \vsix technologies.

Alleviating the privacy and tracking concerns we demonstrate arise from \eui
addressing CPE has a straightforward solution -- \ac{CPE} vendors \emph{must}
ensure their OSes support \ac{SLAAC} privacy extensions and enable their 
use
by default. If supporting privacy extensions is technically infeasible or
undesirable, users must be warned about their increased vulnerability to
tracking by their service provider or equipment vendor. 
Service providers must also take an active role in certifying that the
products they supply to customers meet these privacy and anti-tracking requirements.  Customers have
the right to expect that their privacy will be protected using
state-of-the-art, privacy-enhancing addressing schemes, rather than
jeopardized by CPE employing legacy standards.

\section*{Acknowledgements}
We thank Will van Gulik and Young Hyun for providing measurement 
infrastructure, Jan Sch{\"o}llhammer from AVM GmbH for his responsiveness to our
vulnerability disclosure, and the anonymous reviewers for feedback. Kirstin
Thordarson's thesis~\cite{kirstin} provided valuable early insight for this
study.
This work was supported in part by NSF grants CNS-1855614 and CNS-1901517.  
Views and
conclusions are those of the authors and should not be interpreted as
representing the official policies or position of the U.S.\ government
or the NSF.

\section*{Appendix: Algorithms}
\label{sec:alg}

\begin{algorithm}[!h]
  \caption{Allocation\_Size($AS$)}
\label{alg:allocPrefixSize}
\begin{algorithmic}
  \Require{<periphery response, target address> map $M$ for AS $A$}
  \Ensure{Inferred allocation size}
  \State $eui \leftarrow []$
  \For {$r \in responses$}
    \If {isEUI($r$)}
    \State {$e \leftarrow $ extractEUI($r$)}
    \State {$eui[e] = M[r]$}
    \EndIf
  \EndFor
  \State $sizes \leftarrow []$
  \For {$e \in eui$}
    \State {$min\_r \leftarrow (min(eui[e]) \gg 64)$}
    \State {$max\_r \leftarrow (max(eui[e]) \gg 64)$}
    \State {$size \leftarrow log_2(max\_r - min\_r)$}
    \State {$sizes \leftarrow size \bigcup sizes$}
  \EndFor
  \Return {$median(sizes)$}
\end{algorithmic}
\end{algorithm}

\begin{algorithm}[!h]
  \caption{Rotation\_Pool\_Size($AS$)}
\label{alg:rotationPoolSize}
\begin{algorithmic}
  \Require{Periphery responses for AS $A$}
  \Ensure{Rotation pool size}
  \State $eui \leftarrow []$
  \For {$r \in responses$}
    \If {isEUI($r$)}
    \State {$e \leftarrow $ extractEUI($r$)}
    \State {$eui[e] = r$}
    \EndIf
  \EndFor
  \State $sizes \leftarrow []$
  \For {$e \in eui$}
    \State {$min\_r \leftarrow (min(eui[e]) \gg 64)$}
    \State {$max\_r \leftarrow (max(eui[e]) \gg 64)$}
    \State {$size \leftarrow log_2(max\_r - min\_r)$}
    \State {$sizes \leftarrow size \bigcup sizes$}
  \EndFor
  \Return {$median(sizes)$}
\end{algorithmic}
\end{algorithm}

\begin{acronym}
  \acro{AS}{Autonomous System}
  \acro{ASN}{\ac{AS} Number}
  \acro{BGP}{Border Gateway Protocol}
  \acro{CPE}{Customer Premises Equipment}
  \acro{DAD}{Duplicate Address Detection}
  \acro{EUI}{Extended Unique Identifier}
  \acro{ISP}{Internet Service Provider}
  \acro{IID}{Interface Identifier}
  \acro{LAN}{Local Area Network}
  \acro{NIC}{Network Interface Card}
  \acro{MAC}{Media Access Control}
  \acro{OS}{Operating System}
  \acro{OUI}{Organizationally Unique Identifier}
  \acro{SOHO}{Small Office-Home Office}
  \acro{U/L}{Universal/Local}
  \acro{SLAAC}{Stateless Address Autoconfiguration}
\end{acronym}


\begin{thebibliography}{10}

\bibitem{ieee802}
{IEEE Standard for Local and Metropolitan Area Networks: Overview and
  Architecture}.
\newblock {\em {IEEE Std 802-2014 (Revision to IEEE Std 802-2001)}}, pages
  1--74, 2014.

\bibitem{avm}
{AVM}, 2020.
\newblock \url{https://en.avm.de/}.

\bibitem{bihnet}
{BH Telecom}, 2020.
\newblock \url{https://www.bhtelecom.ba/}.

\bibitem{entel}
{Entel Bolivia}, 2020.
\newblock \url{https://www.entel.bo/}.

\bibitem{oui}
{IEEE OUI database}, 2020.
\newblock \url{http://standards-oui.ieee.org/oui.txt}.

\bibitem{starcat}
{Starcat Cable Network}, 2020.
\newblock \url{http://www.starcat.co.jp.e.lh.hp.transer.com/}.

\bibitem{imc16yarrp}
Robert Beverly.
\newblock {Yarrp'ing the Internet: Randomized High-Speed Active Topology
  Discovery}.
\newblock In {\em Proceedings of ACM Internet Measurement Conference (IMC)},
  November 2016.

\bibitem{imc18beholder}
Robert Beverly, Ramakrishnan Durairajan, David Plonka, and Justin~P. Rohrer.
\newblock {In the IP of the Beholder: Strategies for Active IPv6 Topology
  Discovery}.
\newblock In {\em Proceedings of ACM Internet Measurement Conference (IMC)},
  November 2018.

\bibitem{caida-routed48}
{CAIDA}.
\newblock {The CAIDA UCSD IPv6 Routed /48 Topology Dataset}, 2019.
\newblock
  \url{https://www.caida.org/data/active/ipv6\_routed\_48\_topology\_dataset.xml}.

\bibitem{rfc4443}
A.~Conta, S.~Deering, and M.~{Gupta (Ed.)}.
\newblock {Internet Control Message Protocol (ICMPv6) for the Internet Protocol
  Version 6 (IPv6) Specification}.
\newblock RFC 4443 (Internet Standard), March 2006.
\newblock Updated by RFC 4884.

\bibitem{rfc7721}
A.~Cooper, F.~Gont, and D.~Thaler.
\newblock {Security and Privacy Considerations for IPv6 Address Generation
  Mechanisms}.
\newblock RFC 7721 (Informational), March 2016.

\bibitem{durumeric2013zmap}
Zakir Durumeric, Eric Wustrow, and J~Alex Halderman.
\newblock Zmap: Fast internet-wide scanning and its security applications.
\newblock In {\em {22nd USENIX Security Symposium (USENIX Security 13)}}, pages
  605--620, 2013.

\bibitem{Gasser:2018:CEU:3278532.3278564}
Oliver Gasser, Quirin Scheitle, Pawel Foremski, Qasim Lone, Maciej
  Korczy\'{n}ski, Stephen~D. Strowes, Luuk Hendriks, and Georg Carle.
\newblock {Clusters in the Expanse: Understanding and Unbiasing IPv6 Hitlists}.
\newblock In {\em Proceedings of ACM Internet Measurement Conference (IMC)},
  2018.

\bibitem{anonymousemail}
AVM GmbH.
\newblock {EUI-64 Vulnerability Disclosure}.

\bibitem{rfc7707}
F.~Gont and T.~Chown.
\newblock {Network Reconnaissance in IPv6 Networks}.
\newblock RFC 7707 (Informational), March 2016.

\bibitem{rfc3177}
IAB and IESG.
\newblock {Recommendations on IPv6 Address Allocations to Sites}.
\newblock RFC 3177 (Informational), September 2001.

\bibitem{imc20facebook}
Frank Li and David Freeman.
\newblock {Towards A User-Level Understanding of IPv6 Behavior}.
\newblock In {\em Proceedings of ACM Internet Measurement Conference (IMC)},
  October 2020.

\bibitem{rfc4429}
N.~Moore.
\newblock {Optimistic Duplicate Address Detection (DAD) for IPv6}.
\newblock RFC 4429, April 2006.

\bibitem{rfc8415}
T.~Mrugalski, M.~Siodelski, B.~Volz, A.~Yourtchenko, M.~Richardson, S.~Jiang,
  T.~Lemon, and T.~Winters.
\newblock {Dynamic Host Configuration Protocol for IPv6 (DHCPv6)}.
\newblock RFC 8415 (Proposed Standard), November 2018.

\bibitem{rfc4941}
T.~Narten, R.~Draves, and S.~Krishnan.
\newblock {Privacy Extensions for Stateless Address Autoconfiguration in IPv6}.
\newblock RFC 4941 (Draft Standard), September 2007.

\bibitem{rfc6177}
T.~Narten, G.~Huston, and L.~Roberts.
\newblock {IPv6 Address Assignment to End Sites}.
\newblock RFC 6177 (Best Current Practice), March 2011.

\bibitem{atlas}
RIPE NCC.
\newblock {RIPE Atlas}, 2021.
\newblock \url{https://atlas.ripe.net/}.

\bibitem{akamaiblog}
Erik Nygren.
\newblock {At 21Tbps, Reaching New Levels of IPv6 Traffic}, 2020.
\newblock
  \url{https://blogs.akamai.com/2020/02/at-21-tbps-reaching-new-levels-of-ipv6-traffic.html}.

\bibitem{padmanabhan2020dynamips}
Ramakrishna Padmanabhan, John~P Rula, Philipp Richter, Stephen~D Strowes, and
  Alberto Dainotti.
\newblock {DynamIPs: Analyzing Address Assignment Practices in IPv4 and IPv6}.
\newblock In {\em Proceedings of the 16th International Conference on emerging
  Networking EXperiments and Technologies}, pages 55--70, 2020.

\bibitem{Plonka:2015:TSC:2815675.2815678}
David Plonka and Arthur Berger.
\newblock {Temporal and Spatial Classification of Active IPv6 Addresses}.
\newblock In {\em Proceedings of ACM Internet Measurement Conference (IMC)},
  2015.

\bibitem{rv}
{Routeviews}.
\newblock {University of Oregon Route Views Project}, 2020.
\newblock \url{http://www.routeviews.org/routeviews/}.

\bibitem{edgy}
Erik~C Rye and Robert Beverly.
\newblock {Discovering the IPv6 Network Periphery}.
\newblock In {\em International Conference on Passive and Active Network
  Measurement}, pages 3--18. Springer, 2020.

\bibitem{rfc4862}
S.~Thomson, T.~Narten, and T.~Jinmei.
\newblock {IPv6 Stateless Address Autoconfiguration}.
\newblock RFC 4862, September 2007.

\bibitem{kirstin}
Kirstin~E Thordarson.
\newblock {Analysis of EUI-64-Based Addressing and Associated Vulnerabilities}.
\newblock Master's thesis, Monterey, CA; Naval Postgraduate School, 2020.

\bibitem{zmap6}
tumi8.
\newblock {ZMapv6: Internet Scanner with IPv6 Capabilities}, 2021.
\newblock \url{https://github.com/tumi8/zmap}.

\end{thebibliography}
\end{document}